\documentclass[final,5p,times, twocolumn]{elsarticle}
\usepackage{amssymb} % various useful mathematical symbols
\usepackage{amsmath} % various useful equation environments
\usepackage[style=apa,sorting=none]{biblatex} % bibliography
\usepackage{booktabs} % pretty tables
\usepackage{etoolbox}
\usepackage{float}
\usepackage[acronym, nonumberlist]{glossaries}
\usepackage{graphicx} % table auto-fitting
\usepackage[switch]{lineno} % outer margins in two-column
\usepackage{mdframed}  % simpleboxes
\usepackage{nomencl}  % aid for math definitions
\usepackage{subcaption} % Multi-figure plots
\usepackage{svg}  % High-quality SVG images
\usepackage[colorlinks,allcolors=blue]{hyperref}  % clickable hyperlinks
\usepackage{cleveref}  % better referencing

\makenomenclature
\makeglossaries

\newacronym{capex}{CAPEX}{capacity expenditures}
\newacronym{opex}{OPEX}{operational expenditures}
\newacronym{eia}{EIA}{U.S. Energy Information Administration}
\newacronym{esom}{ESOM}{Energy System Optimisation Model}
\newacronym{ghg}{GHG}{greenhouse gas}
\newacronym{iam}{IAM}{Integrated Assessment Model}
\newacronym{iea}{IEA}{International Energy Agency}
\newacronym{ipcc}{IPCC}{Intergovernmental Panel on Climate Change}
\newacronym{lp}{LP}{linear programming}
\newacronym{milp}{MILP}{mixed-integer linear programming}
\newacronym{minlp}{MINLP}{mixed-integer non-linear programming}
\newacronym{nlp}{NLP}{non-linear programming}
\newacronym{prisma}{PRISMA}{Preferred Reporting Items for Systematic reviews and Meta-Analyses}
\newacronym{qr}{QR}{quantile regression}
\newacronym{rcp}{RCP}{Representative Concentration Pathway}
\newacronym{ssp}{SSP}{Shared Socioeconomic Pathway}
\newacronym{td-idf}{TD-IDF}{Term Frequency-Inverse Document Frequency}
\newacronym{weo}{WOE}{World Energy Outlook}

\renewcommand\nomgroup[1]{%
  \item[\bfseries
  \ifstrequal{#1}{A}{Index sets}{%
  \ifstrequal{#1}{B}{Parameters}{%
  \ifstrequal{#1}{C}{Variables}{%
  \ifstrequal{#1}{D}{Equations}{}}}}%
]}

\nomenclature[A]{$h \in H$}{Hours}
\nomenclature[A]{$m \in M$}{Investment milestones}
\nomenclature[A]{$r \in R$}{Spatial regions}
\nomenclature[A]{$t \in T$}{Operational timesteps}
\nomenclature[A]{$y \in Y$}{Years}
\nomenclature[A]{$i \in I$}{Investment options}

\nomenclature[B]{$\mathbf{ML}_{m}$}{Milestone length}
\nomenclature[B]{$\mathbf{d}$}{Discount rate}
\nomenclature[B]{$\mathbf{IC}_{im}$}{Investment cost}
\nomenclature[B]{$\mathbf{Life}_i$}{Lifetime of an investment}
\nomenclature[B]{$\mathbf{LifeRem}_{im}$}{Remaining lifetime of an investment}
\nomenclature[B]{$\mathbf{WACC}_i$}{Weighted average cost of capital}

\nomenclature[C]{$nc_{irm}$}{New capacity}

\nomenclature[D]{$CAPEX_m$}{Capacity Expenditures}
\nomenclature[D]{$OPEX_m$}{Operational Expenditures}
\nomenclature[D]{$AF_{i}$}{Annuity factor}
\nomenclature[D]{$SV_{imr}$}{Salvage value}

\begin{document}

\begin{frontmatter}

%% Title, authors and addresses

%% use the tnoteref command within \title for footnotes;
%% use the tnotetext command for theassociated footnote;
%% use the fnref command within \author or \affiliation for footnotes;
%% use the fntext command for theassociated footnote;
%% use the corref command within \author for corresponding author footnotes;
%% use the cortext command for theassociated footnote;
%% use the ead command for the email address,
%% and the form \ead[url] for the home page:
%% \title{Title\tnoteref{label1}}
%% \tnotetext[label1]{}
%% \author{Name\corref{cor1}\fnref{label2}}
%% \ead{email address}
%% \ead[url]{home page}
%% \fntext[label2]{}
%% \cortext[cor1]{}
%% \affiliation{organization={},
%%             addressline={},
%%             city={},
%%             postcode={},
%%             state={},
%%             country={}}
%% \fntext[label3]{}

\title{Optimising for the long game: methodological challenges in energy system optimisation pathways}

%% use optional labels to link authors explicitly to addresses:
%% \author[label1,label2]{}
%% \affiliation[label1]{organization={},
%%             addressline={},
%%             city={},
%%             postcode={},
%%             state={},
%%             country={}}
%%
%% \affiliation[label2]{organization={},
%%             addressline={},
%%             city={},
%%             postcode={},
%%             state={},
%%             country={}}

\author[tudelft]{Ivan Ruiz Manuel} %% Author name
\ead{i.ruizmanuel@tudelft.nl}
\author[tudelft]{Meijun Chen} %% Author name
\author[tudelft]{Francesco Lombardi} %% Author name
\author[tudelft]{Stefan Pfenninger-Lee} %% Author name

%% Author affiliation
\affiliation[tudelft]{
    organization={TU Delft, Faculty of Technology, Policy and Management, Department of Engineering Systems and Services},%Department and Organization
    addressline={Jaffalaan 5},
    city={Delft},
    postcode={2628 BX},
    state={South Holland},
    country={Netherlands}
}

%% Abstract
\begin{abstract}
%% Text of abstract
Pathways that describe the optimal evolution of energy systems across multiple decades are important in energy system research and policy literature, with net-zero and similar climate policies being common drivers behind them. While there are many studies on aspects such as spatial and operational resolution, model features, and model transparency, there has been little attention on the methodological considerations of formulating pathway studies in mathematical optimisation terms, and how these methods have evolved over time. To address this, we conduct a systematic review of optimal pathway literature at or above national level focusing on the following: i) the implications of model foresight choices, ii) end effects and related issues that may bias model outcomes, iii) trade-offs in model resolution, and iv) investment dynamics. We showcase how modellers have dealt with these aspects in a large sample of studies spanning multiple decades, and provide recommendations to both modellers and model users on identifying issues that can bias model results and how to improve upon them. In particular, we identify opportunities to better balance long-term anticipatory planning with high operational and spatial detail in models, and to improve the communication and systematic treatment of those mathematical design choices that potentially distort model decisions across time.
\end{abstract}

% %% Graphical abstract
% \begin{graphicalabstract}
% %\includegraphics{grabs}
% \end{graphicalabstract}

%%Research highlights
% \onecolumn
% \begin{highlights}
% \item Quantitative assessment of \acrlong{esom} literature on long-term pathways published over recent decades.
% \item Theoretical synthesis of critical aspects to consider when evaluating pathway exercises.
% \item Evidence of a shift towards near-sighted model setups, driven by increased use of myopic approaches and shorter model horizons.
% \item Models have achieved finer operational and spatial detail, but long-term aspects remain coarse.
% \item Better practices could reduce model distortions and misinterpretation of pathway model results.
% \end{highlights}
% \twocolumn

%% Keywords
\begin{keyword}
%% keywords here, in the form: keyword \sep keyword
Energy system modelling \sep Optimisation \sep Capacity expansion \sep Pathways \sep Systematic Literature Review
%% PACS codes here, in the form: \PACS code \sep code

%% MSC codes here, in the form: \MSC code \sep code
%% or \MSC[2008] code \sep code (2000 is the default)

\end{keyword}

\end{frontmatter}

\printnomenclature

% \linenumbers
\begin{refsection}
\section{Introduction}
\label{sec:intro}
% Scaffold:
% - Discuss what pathways are, how they differ from regular scenarios and why they are prominent in sustainability / energy literature.
% - Detail how ESOMs relate to pathways and the unique characteristics that these tools bring to pathway creation, as well as their advantages and disadvantages. Bring out the optimisation of carbon budgets as a case. Bring out how optimisation models are at the heart of many C-IAMs
% - Give a brief summary of reviews and meta-studies done in relation to ESOMs and pathways and the gaps left by them.
% - Bring out the topic of study / model transparency.
% - Detail the scope of this review and how it compliments previous research (the "gap" filled).
% - Obligatory summary of each section.
Pathways, i.e. trajectories that describe the plausible evolution of a system over a period of years to decades, are important means of planning and assessing changes in techno-economic and socio-technical systems, particularly in the context of mitigating climate change~\parencite{rosenbloom_pathways_2017, turnheim_evaluating_2015}. The concept of ``pathways'' is often used interchangeably with that of ''scenarios``~\parencite{maier_uncertain_2016, swart_problem_2004}, although the latter can also be used to describe the end state of a possible future without necessarily detailing the developments that lead to it~\parencite{borjeson_scenario_2006}. Pathways can be produced qualitatively by using narratives to describe the interactions between actors and the systems surrounding them~\parencite{geels_typology_2007}, or quantitatively with the aid of computer models in combination with the aforementioned narratives~\parencite{turnheim_evaluating_2015}.

The transition of energy systems to climate neutrality has been at the centre of many quantitative pathway exercises due to the large contribution of such systems to \gls{ghg} emissions~\parencite{ipcc_climate_2023}. Given the highly complex nature of the energy transition, pathways detailing energy investments several decades into the future have become ubiquitous, with governments and international bodies commissioning them as part of policy processes~\parencite{susser_model-based_2021}. Examples of pathway exercises widely featured in research and policy documents include the \glspl{ssp}~\parencite{oneill_new_2014} and \glspl{rcp}~\parencite{van_vuuren_representative_2011} produced as part of \gls{ipcc} reports, the \gls{iea}['s] annually published \acrlong{weo}~\parencite{international_energy_agency_world_2024} and the \gls{eia}['s] International Energy Outlook~\parencite{us_energy_information_administration_international_2023}.

One of the most commonly used methods to produce energy system pathways are bottom-up \glspl{esom}, with the aim of minimising monetary investment and operation costs from a central planner perspective~\parencite{pfenninger_energy_2014, shu_overcoming_2024}. This type of goal-oriented ---i.e., ``normative''--- pathway has a storied history in governmental and electric utility planning~\parencite{hoffman_energy_1976}, with models such as TIMES~\parencite{loulou_documentation_2016}, OSeMOSYS~\parencite{howells_osemosys_2011} and MESSAGE~\parencite{huppmann_messageix_2019} being widely recognised in scientific literature. The approach has also influenced international climate policy, as many of the \glspl{iam} used to investigate the long-term evolution of \gls{ghg} emissions have an \gls{esom} at their core~\parencite{pietzcker_system_2017}.

The application of \glspl{esom} to highly uncertain long-term questions has been subjected to a range of criticisms, including how their insufficient spatio-temporal resolution biases results against renewable technologies~\parencite{pfenninger_energy_2014}, their frequent use of highly subjective parameters and formulations~\parencite{ellenbeck_how_2019}, their use of inaccurate or outdated cost projections of renewable technologies~\parencite{jaxa-rozen_sources_2021, creutzig_underestimated_2017}, and a general lack of transparency that obfuscates the interpretation of their results~\parencite{pfenninger_importance_2017, decarolis_case_2012}. Energy modelling literature has responded to said criticisms by, for example, developing and assessing novel techniques for time series aggregation to better represent the patterns of demand and renewable resources~\parencite{kotzur_time_2018, pfenninger_dealing_2017}, by comparing the effects of different model formulations~\parencite{candas_code_2022} or classifying models by their scope or features~\parencite{plazas-nino_national_2022, lopion_review_2018, prina_classification_2020}.

Despite all this attention, the methods and particularities of pathways produced by \gls{esom} are seldom discussed with the same degree of detail. Nevertheless, producing quantitative pathways is one of the most important applications of such models, and a model's long-term setup can affect outcomes just as much (or more) as the issues discussed above. For example, assessments of energy system models have noticed a general lack of discourse on the topic of model foresight and its implications~\parencite{ellenbeck_how_2019}, which is reflected in how reviews of energy models often omit the topic entirely~\parencite{fattahi_systemic_2020, fodstad_next_2022, gacitua_comprehensive_2018, chang_trends_2021} or only discuss it tangentially~\parencite{kotzur_modelers_2021}. Yet, studies assessing this aspect regularly highlight its significant influence on model results~\parencite{heuberger_power_2017, keppo_short_2010}. Similarly, despite the mounting evidence on the importance of models having adequate spatial and temporal resolution~\parencite{pfenninger_dealing_2017, poncelet_impact_2016, prina_classification_2020}, to date there is no systematic analysis of whether models have actually improved in that regard. Other nuances of long-term optimisation model formulations, such as the presence and prevention of end effects~\parencite{mavromatidis_mango_2021,grinold_time_1980} or the implications of different technological learning approaches~\parencite{behrens_reviewing_2024} have garnered even less discussion. To our knowledge, there is no systematic assessment of \gls{esom} literature in regards to how pathway exercises are conducted and the considerations therein, which is worrisome given that this lack of awareness could result in disconnects between the narrative modellers intend to create within their modelling exercises, the ways an optimisation model will behave in the presence of distortions and other modelling artifacts, and how model results may be interpreted by users who lack enough context on the impacts of modeller choices.

Here we conduct a systematic review of pathway studies using \glspl{esom} with a focus on the methodological aspects of these exercises. In particular, we focus on: i) the assumptions and justifications given when using different types of foresight and their implications; ii) the concept of the decision horizon and how model distortions such as end effects can constrain it; iii) tradeoffs in model resolution; and iv) the techniques used by modellers to influence investment dynamics. By describing these aspects in an accessible way, we hope to provide model developers with insights on how to further improve their frameworks, and to aid model users in contextualising pathway results in the face of weaknesses that might be present in the models that produce them. Our focus is on discussing key aspects of pathway exercises and how they have changed over the years, not on evaluating the sectoral coverage of models, the quality of their data, or features that may improve the operational or technological formulations within models. Finally, we limit our focus to \gls{esom} studies at national or supernational scale, in order to focus on model formulation issues that might impact national policy advice, and to ensure a reasonably homogenous sample of studies.

The paper is organised as follows. We start by detailing several methodological considerations of pathway exercises in \cref{sec:theory}, which will serve as a lens to detect potential issues in the current practices emerging from our subsequent literature analysis. \Cref{sec:methods} provides an overview of our systematic review method, including research queries, the protocols and tools utilised, and a description of our approach to data collection. \Cref{sec:results} compiles the results of our analyses, including bibliometric approaches, and the resulting statistics of our sample of studies. Finally, \cref{sec:discussion} concludes with discussions and recommendations for pathway exercises going forward.

\section{Background and theory}
\label{sec:theory}

% This section will mostly be a repeat of the important bits of the pathway techniques and considerations document to contextualise the reader on important aspects we'll focus on.
% It must have the following (in order):

% \begin{enumerate}
%     \item Defining what multi-period / dynamic optimisation is and how time horizons play a part in it. DONE.
%     \item Defining a consistent nomenclature for model foresight. In particular, we should define what static problems are, when myopia / limited foresight can mislead in term of optimal choices. Generic formulations for each type of problem should be given here. DONE.
%     \item Describing the difference between the decision horizon and the model horizon, and why discount rates are necessary when optimising multi-period problems. End-of-horizon distortions should be introduced here, as well as typical approaches to mitigate them. DONE
%     \item Description of how models typically aggregate years in ``milestones'' and their nuances (inter-milestone distortions). We should also clearly state that myopic models using milestones cannot deal with these distortions due to their static nature. DONE
%     \item Discussion of ``opportunistic'' behaviour in these models and how modellers try to deal with it. Important aspects to touch upon are: endogenous learning, capacity / generation growth constraints, and the influence that high discount rates have on model results. DONE
% \end{enumerate}

Constructing an \gls{esom} involves many choices and assumptions by model developers, which in turn affect the questions said model is best suited to answer; these choices and assumptions are also important aspects to consider when interpreting results~\parencite{decarolis_formalizing_2017}. In this section, we detail multiple aspects we have identified as essential when it comes to pathway formulations in \glspl{esom}, placing particular focus on details often ignored in the literature. Our aim is to sufficiently explain these methodological aspects while providing a consistent nomenclature to describe them. These concepts will guide us in our subsequent analysis of the literature, helping us to detect any current practices that may be overlooking important methodological considerations. In particular, we focus on i) describing model horizons and foresight, ii) detailing end effects and how to mitigate them, iii) explaining resolution tradeoffs with a particular focus on how models aggregate investment decisions to improve tractability, and iv) examining how models regulate investment dynamics.

\subsection{Time horizons and foresight}
\label{subsec:model-horizons}

\begin{figure}[bt]
    \centering
    \includegraphics[width=1\linewidth]{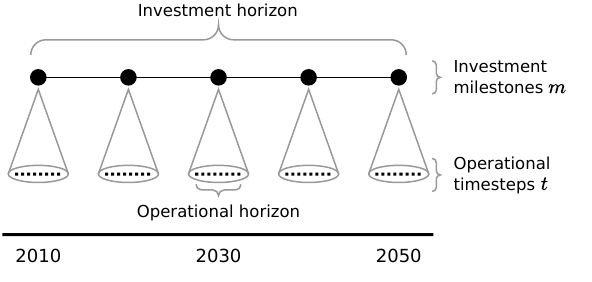}
    \caption{The different horizons considered in a typical \gls{esom} pathway exercise.}
    \label{fig:theory-horizons}
\end{figure}

Possibly the most important concept to keep in mind when evaluating pathways is the time horizon under study and how the model considers it when making decisions. A distinction should be made between \textit{investment horizons}, which are the periods when the model can decide to invest or remove certain technologies, and \textit{operational horizons}, in which the model can only decide how to best utilise the technologies already at its disposal (\Cref{fig:theory-horizons}).

\begin{figure}[bt]
    \centering
    \includegraphics[width=1\linewidth]{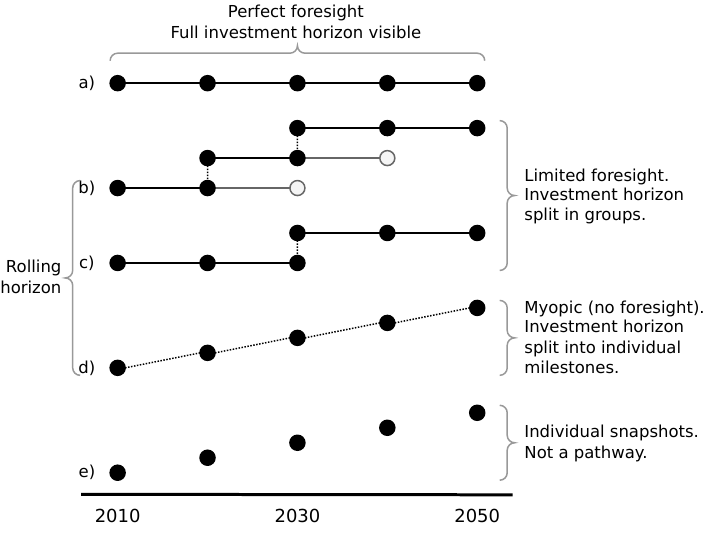}
    \caption{Classification of typical foresight approaches seen in \gls{esom} pathways. Dots represent model milestones, with white dots representing reconsidered investments. Adapted from~\textcite{keppo_short_2010, kotzur_modelers_2021}.}
    \label{fig:theory-foresight}
\end{figure}

Modellers might want to limit a model's ability to anticipate events depending on their research goals and the computational capabilities at hand. For instance, when evaluating how delayed investment and systemic inertia can affect policy goals~\parencite{heuberger_power_2017, mannhardt_understanding_2024}. This means reducing the model's \textit{foresight}---often called the planning or decision horizon~\parencite{grinold_time_1980, keppo_short_2010}---which is the degree of visibility that the model has into future system conditions. A reduced foresight leads the model to obtain suboptimal results within the bounds of the narrative constructed by the modellers. In general, three kinds of foresight approaches can be distinguished  (\Cref{fig:theory-foresight}).

Under \textit{perfect foresight} the model has full visibility of the investment horizon, including price developments, future policies, and decommissioning requirements. It can be thought of as the most normative approach~\parencite{borjeson_scenario_2006}, since it is the only one that guarantees optimality within the bounds of the problem. However, perfect anticipation over multiple decades does not reflect reality, meaning this type of exercise is ill-suited to reproduce actual policymaker behaviour~\parencite{fuso_nerini_myopic_2017} and is often overly optimistic when exploring system disruptions~\parencite{heuberger_power_2017}. A simplified version of most perfect foresight formulations is shown in \cref{eq:perfect-obj}, where the model minimises capacity and operational expenditures ($\mathit{CAPEX}_m$ and $\mathit{OPEX}_m$, respectively) for all the milestone years under consideration ($m$):

\begin{equation} \label{eq:perfect-obj}
    \mathrm{Min :} \sum_{m} \mathbf{DF}_m(CAPEX_{m}+ OPEX_{m})
\end{equation}

Rolling horizon approaches mitigate an \gls{esom}'s anticipatory behaviour by subdividing the investment horizon into groups and then passing---or ``rolling''---results from one group to the next~\parencite{kotzur_modelers_2021}. We can subdivide this category into two distinct methods: \textit{limited foresight} and \textit{myopic foresight}. Under limited foresight, models retain some anticipatory ability and may even be allowed some overlap between groups of decisions to enable reconsideration in the face of new information~\parencite{keppo_short_2010, mannhardt_understanding_2024}. This means they are in essence a sequence of connected perfect foresight problems. In contrast, fully myopic models essentially lose all anticipatory ability, optimising only single investment periods. In essence, such models assume that agents do not know about or actively decide to ignore expected developments and instead face them as they occur~\parencite{kydes_beyond_1995}. \Cref{eq:myopic-obj} shows a simplified equivalent of a myopic model:

\begin{equation} \label{eq:myopic-obj}
    \mathrm{Min :} \ CAPEX_{m}+ OPEX_{m}
\end{equation}

Although myopic models are sometimes seen as more realistic~\parencite{martinsen_implications_2007}, one should keep in mind that these models often retain perfect anticipation for their operational decisions (e.g., \textcite{abuzayed_mypypsa-ger_2022, plesmann_how_2017}). Whether or not forgoing anticipatory investment is a desired or realistic characteristic depends on the research question, as studies comparing the effects of foresight in ESOMs have highlighted both the dangers of using perfect foresight models with unrealistic ``unicorn'' expectations~\parencite{heuberger_power_2017}, and how myopic planning poses risks of locking-in undesirable system states~\parencite{mannhardt_understanding_2024}.

\subsection{End effects and decision horizons}
\label{subsec:end-effects}

Another aspect complicating pathway exercises is the finite horizon of the optimisation itself and the distortions it brings. Energy system pathways, like all capacity expansion formulations, are a type of infinite horizon problem because the period in which one needs to install and decommission investments is never-ending~\parencite{bean_conditions_1984}. Since solving these models over an infinite horizon is intractable and would involve increasingly uncertain assumptions, modellers are forced to choose a finite period to focus on instead~\parencite{smith_planning_1981, grinold_model_1983, krishnan_building_2017}. Energy modellers rarely discuss the implications of this choice even though operations research literature has long acknowledged its importance~\parencite{chand_forecast_2002}.

\begin{figure}[bt]
    \centering
    \includegraphics[width=1\linewidth]{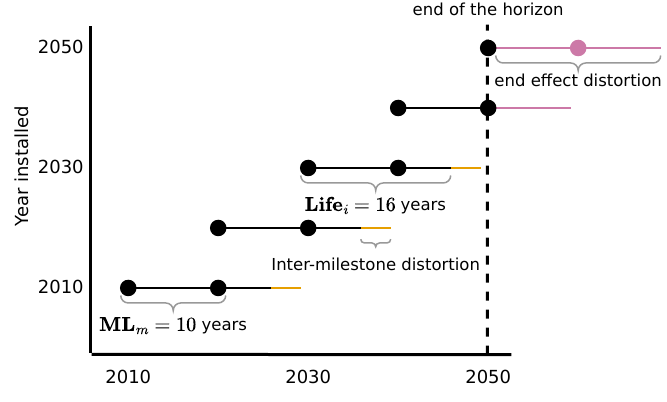}
    \caption{Simplified example of common model distortions caused by interactions between the lifetime of investments ($\mathbf{Life}_i$), foresight horizon (40 years), and milestone length ($\mathbf{ML}_m$) in a model with perfect foresight.}
    \label{fig:model-distortions}
\end{figure}

% decision horizon, then discounting then end of horizon effects
Making an infinite horizon problem finite results in what we can call \textit{end effects}, which are distortions that cause the model to make increasingly short-sighted choices the closer it is to the end of its foresight horizon~\parencite{bean_optimal_1985}. In the case of \glspl{esom}, these will typically impact investment decisions due to improper consideration of investment lifetimes ($\mathbf{Life}_i$) versus their operational cost (\Cref{fig:model-distortions}), often artificially benefiting options with low initial investment cost but high operational cost such as natural gas turbines~\parencite{grinold_time_1980} or short-term decarbonisation solutions such as co-firing and carbon capture and storage retrofits~\parencite{krishnan_building_2017}. In turn, they negatively affect investments with higher installation cost but low operational expenses (e.g., solar photovoltaics or wind turbines), and investments with long lifetimes such as hydropower~\parencite{krishnan_building_2017}. 

The longer a model's foresight horizon is, the less its end effects will impact earlier decisions~\parencite{chand_forecast_2002}, meaning that it is necessary to extend the horizon of a model beyond the target year of interest in order to avoid distorted assessments~\parencite{bean_optimal_1985}. Operations research often refers to the \textit{decision horizon} as the minimum length of the foresight horizon that guarantees that the first choice of the model (i.e., the first investment year in the case of an \gls{esom}) is unaffected by end effects~\parencite{ghate_infinite_2011}. This can make \glspl{esom} intractable since they often deal with long-term targets and assets with long lifetimes~\parencite{grinold_time_1980}. Modellers often turn to alternatives to mitigate end effects, and thus reduce the decision horizon, in two ways: by discounting and by adjusting investment cost calculations~\parencite{krishnan_building_2017}.

Discounting implies using a discount factor ($\mathbf{DF}_m$) to diminish the present value of future costs based on an assumed social discount rate (\cref{eq:perfect-obj,eq:discounting}), with values between 2-10\% being commonly seen in the literature~\parencite{garcia-gusano_role_2016}, and a typical formulation for $\mathbf{DF}_m$ as follows:

\begin{equation}
    \mathbf{DF}_m = \frac{1}{(1+\mathbf{d})^{m-m_0}} \quad \forall \ m\in M
    \label{eq:discounting}
\end{equation}

\begin{figure}[bt]
    \centering
    \includegraphics[width=1\linewidth]{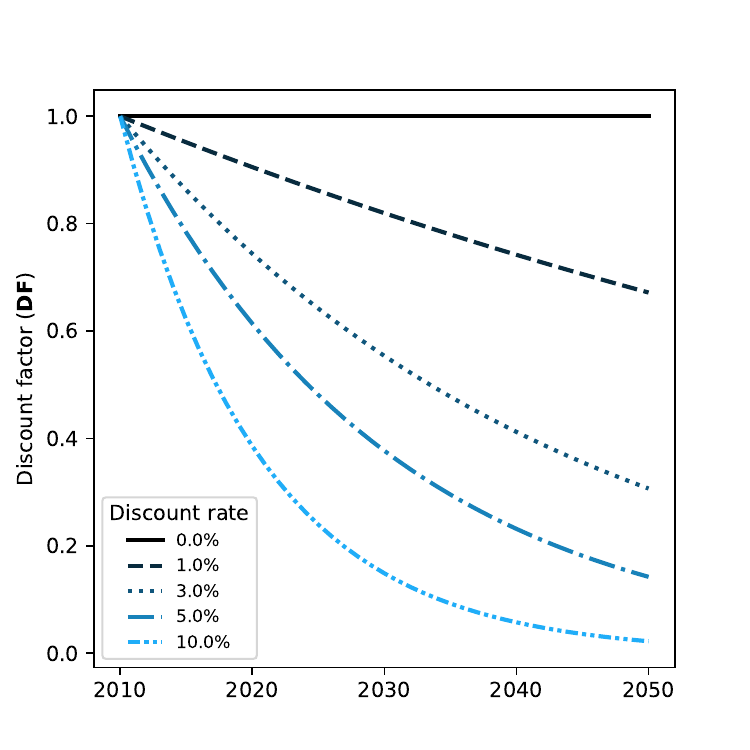}
    \caption{Discount factor ($\mathbf{DF}_m$) trends at different discount rates, relative to an assumed initial model year of 2010. For illustration, a €1,000 investment cost incurred in 2050 corresponds to €671, €142, and €22 in 2010-equivalent terms under discount rates of 1\%, 5\%, and 10\%, respectively.}
    \label{fig:discount-factor}
\end{figure}

The choice of the key parameter in this, the discount rate ($\mathbf{d}$), is complex and value-laden. From a mathematical perspective, higher discount rates will reduce the decision horizon required to make an initial choice by attenuating later costs relative to earlier ones (see \Cref{fig:discount-dynamics}). However, discounting implies intergenerational inequities which are particularly important to consider in the context of climate change~\parencite{dasgupta_discounting_2008, pindyck_climate_2013}. \glspl{esom} are particularly sensitive to this parameter, with even small increments in the discount rate leading to significant changes in the discount factor between investment milestones, and thus in the options selected by the model (see \Cref{fig:discount-factor}). Higher discount rates often delay the implementation of decarbonisation technologies~\parencite{garcia-gusano_role_2016}, leading to calls to use lower values~\parencite{loffler_social_2021}, with some modellers even choosing to use no discounting at all~\parencite{zeyen_endogenous_2023}.

To reduce artificial disadvantages affecting investments near the end of the horizon---both relative to earlier investments and between late investments themselves---discounting is often complemented by adjustments to investment cost calculations. The most common approaches are salvage values and annuity factors, which influence the timing at which the model accounts for technology investment costs~\parencite{candas_code_2022, brown_multi-horizon_2020}. An \gls{esom} without these adjustments will typically estimate total capacity expenditures in a given year ($CAPEX_m$) by multiplying the total newly installed capacity ($nc_{imr}$) by its overnight installation cost ($\mathbf{IC}_{im}$), as shown in \cref{eq:capex}:

\begin{equation}
    CAPEX_m =  \sum_{i,r}  \mathbf{IC}_{im}\,nc_{imr} \quad \forall \ m\in M
    \label{eq:capex}
\end{equation}

In this formulation, technologies closer to the end of the horizon will require installation expenses in full while seeing reduced use, leading to distorted estimates~\parencite{mavromatidis_mango_2021}. Studies evaluating end effects often call this method ``truncation''~\parencite{grinold_time_1980, krishnan_building_2017} and, while problematic, it is not uncommon to see it in the literature (e.g., \textcite{heuberger_power_2017,trutnevyte_does_2016,guerra_optimization_2016}).

The first adjustment method, \textit{salvage values}, adjusts \cref{eq:capex} by subtracting a residual for investments with a remaining lifetime at the end of the horizon ($SV_{imr}$). These residual values attempt to represent the expected benefits that were not captured within the foresight horizon, and are used by popular modelling frameworks such as TIMES~\parencite{loulou_documentation_2016} and OSeMOSYS~\parencite{howells_osemosys_2011}. \Cref{eq:capex-sv,eq:sv} show a simplified example of this method based on~\textcite{howells_osemosys_2011}:

\begin{equation}
    \begin{split}
        CAPEX_m^{SV} = \sum_{i,r} \left( \mathbf{IC}_{imr}\,nc_{imr} - SV_{imr}\right)\\
        \quad \forall \ m\in M
    \end{split}
    \label{eq:capex-sv}
\end{equation}
\begin{equation}
    \begin{split}
    SV_{imr} = 
        \begin{cases}
            \text{if } m+\mathbf{Life}_{i}-1 \leq m_{max}\text{:}\ 0,\\
            \text{else:}\ \mathbf{IC}_{im}\ nc_{imr}\left(1-\dfrac{(1+\mathbf{d})^{m_{max}-m+1}-1}{(1+\mathbf{d})^{\mathbf{Life}_{i}}-1}\right)
        \end{cases}\\
    \forall \ i\in I, m\in M, r \in R
    \end{split}
    \label{eq:sv}
\end{equation}

The second adjustment method, \textit{annuity factors} ($AF_i$), subdivides capacity costs into annuitised yearly payments, often adjusted by some investment-specific weighted average cost of capital ($\mathbf{WACC}_i$)~\parencite{candas_code_2022, hunter_modeling_2013}. Technologies are compared only by accounting for the payments up to the end-of-horizon. This formulation is also popular, with modelling frameworks such as TEMOA~\parencite{hunter_modeling_2013} and PyPSA~\parencite{zeyen_endogenous_2023} making use of it. \Cref{eq:capex-af,eq:af,eq:life-eoh} show an example of this method based on~\textcite{hunter_modeling_2013}:

\begin{equation} 
    CAPEX_m^{AF} = \sum_{i,r} \left(AF_{i}\,\mathbf{IC}_{imr}\,nc_{imr} \right) \quad \forall m \in M
    \label{eq:capex-af}
\end{equation}
\begin{equation}
    AF_{i} = \frac{\mathbf{WACC}_{i}}{1-(1+\mathbf{WACC}_i)^{\mathbf{LifeEoH}_{im}}} \quad \forall i \in I
    \label{eq:af}
\end{equation}
\begin{equation}
    \begin{split}
        \mathbf{LifeEoH}_{im} =
        \begin{cases}
            \mathbf{Life}_i \ &\text{if} \ m+ \mathbf{Life}_i - 1 \leq m_{max} \\
            m_{max} - m + 1 &\text{otherwise}
        \end{cases}\\
        \forall \ i\in I, m\in M
    \end{split}
    \label{eq:life-eoh}
\end{equation}

Even though the use of discounting, salvage values or annuity factors can help in mitigating end effects, they do not eliminate them. \gls{esom} literature is scarce in evaluating the impacts of these formulations on end effects, but the few studies that exist confirm that end effects are still present even when both discounting and salvage values are used~\parencite{grinold_time_1980, grinold_model_1983}, and highlight that annuity factors are likely to suffer from similar issues~\parencite{krishnan_building_2017}. It is also important to keep in mind that investment lifetimes are just one aspect of end effects. Generally, the more complex a model's long-term formulation is (e.g., decreasing technology efficiencies over time, long-term storage, constraints endogenously affecting technology costs or carbon budget allocation, etc.), the more likely it is that modellers will need to extend the model's foresight beyond the target year of interest in order to avoid distorted results.

\subsection{Spatial and temporal aggregation}
\label{subsec:theory:aggregation}

% general description of the topic, briefly introduce the concept of operational resolution
Using an \gls{esom} is often a balancing act between the computational power at hand and the model's resolution, and modellers often turn to heuristics or clustering techniques to improve tractability. It is common to see models aggregating investment decisions to only a few milestone years ($M$) within the full horizon under study ($Y$), and a reduced subset of timesteps ($T$) within within the hours of each year ($H$) to keep an exercise tractable~\parencite{hunter_modeling_2013, pfenninger_dealing_2017}, as shown in \cref{eq:milestone-set,eq:timestep-set}:

\begin{equation} \label{eq:milestone-set}
    m \in M \subseteq Y
\end{equation}
\begin{equation} \label{eq:timestep-set}
    t \in T \subseteq H
\end{equation}

Doing so comes at a cost in accuracy. The literature has long argued that overly coarse models distort the contribution of renewable technologies~\parencite{pfenninger_energy_2014}. Accordingly, operational time series aggregation has received substantial attention, with studies highlighting how distortions induced by overly coarse time-series data can be significant~\parencite{hoffmann_pareto-optimal_2022, pfenninger_dealing_2017}. Studies exploring spatial aggregation have reached similar conclusions when it comes to investments in renewable technologies and transmission networks~\parencite{siala_impact_2019, frysztacki_inverse_2023}, and have highlighted that higher spatial resolution is always preferred if one wishes to diminish distorting results in relation to costs, siting accuracy, and emission trends~\parencite{jacobson_quantifying_2024}.

The aggregation of milestone years in pathway models has received much less attention, however. The need for it is simple: every year with new investments within the horizon of a model necessitates re-calculation of operational decisions, which can quickly make a model intractable. Similar to other dimensions, aggregating investments implies sacrifices to a model's precision. Accounting for artefacts induced by year aggregation is not simple, and has led to models like TIMES reformulating their approach in order to reduce distortions~\parencite{lehtila_times_2016}.

\begin{figure}[bt]
  \centering
  % Top row (50% + 50%)
  \begin{minipage}[t]{.8\linewidth}
    \centering
    \includegraphics[width=\linewidth]{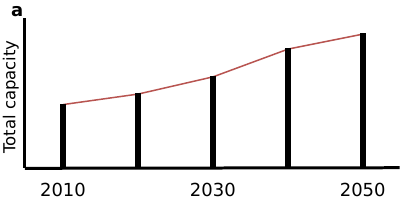}
    \phantomsubcaption\label{fig:capacity-evolution-single}
  \end{minipage}
  \begin{minipage}[t]{.8\linewidth}
    \centering
    \includegraphics[width=\linewidth]{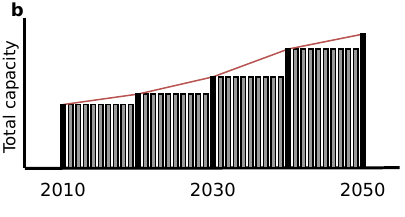}
    \phantomsubcaption\label{fig:capacity-evolution-square}
  \end{minipage}
  \begin{minipage}[t]{.8\linewidth}
    \centering
    \includegraphics[width=\linewidth]{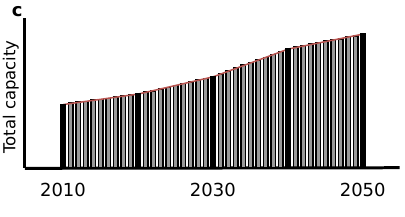}
    \phantomsubcaption\label{fig:capacity-evolution-line}
  \end{minipage}
  \caption{Illustrative example of different milestone interpolation methods seen in the literature. Black bars represent model decisions, grey bars represent interpolated estimations, red line is a linear interpolation case (for comparison). \textbf{a} No interpolation~\parencite{howells_osemosys_2011}. \textbf{b} Constant interpolation~\parencite{hunter_modeling_2013}. \textbf{c} Linear interpolation~\parencite{lehtila_times_2016}.}
  \label{fig:capacity-evolution}
\end{figure}

The methods used by models to interpolate between milestones are diverse (\Cref{fig:capacity-evolution}). Some models, such as OSeMOSYS~\parencite{howells_osemosys_2011}, forgo any kind of interpolation and only calculate costs for the requested milestones (\Cref{fig:capacity-evolution-single}). Others such as TEMOA and PyPSA~\parencite{hunter_modeling_2013, zeyen_endogenous_2023}, assume capacities and generation remain constant between milestones (\Cref{fig:capacity-evolution-square}). Finally, models like TIMES and EFOM possess involved formulations that attempt to distribute new investments equally between milestones~\parencite{european_commission_energy_1984}, with TIMES even offering multiple distinct approaches~\parencite{lehtila_times_2016} (\Cref{fig:capacity-evolution-line}).

Milestone aggregation has important implications in the cost estimations of a model, and may introduce additional distortions to the results, which we call \textit{inter-milestone effects}. In particular, it limits the capacity of a model to adequately assess decommissioning~\parencite{brown_multi-horizon_2020}. If unaccounted for, a model may arbitrarily benefit certain investments with an erroneous extended lifetime if the length between milestones ($\mathbf{ML}_m$) and the remaining lifetime of an asset ($\mathbf{LifeRem}_{im}$) are misaligned (\Cref{fig:model-distortions}, \cref{eq:life-extension-error}).

\begin{equation}
    \begin{split}
        \mathbf{LifeExtErr}_{im} = (-\mathbf{LifeRem}_{im}) \bmod \mathbf{ML}_m\\
        \forall \ i\in I, m\in M
    \end{split}
    \label{eq:life-extension-error}
\end{equation}

Corrections to these distortions vary per modelling framework and interpolation approach. For example, both TEMOA and MESSAGE scale down the distorted capacity by a factor of its remaining lifetime and the length of the distorted milestone: if the milestone length is 10 and the remaining lifetime of an investment is 7, its remaining capacity will be scaled down by 0.7 for that milestone~\parencite{huppmann_messageix_2019, hunter_modeling_2013}. Although this does not remove the distortion---the asset's lifetime is still artificially extended---it changes its nature in an attempt to reduce its influence.

\subsection{Investment dynamics}

When evaluating cost-optimised pathways, one should keep in mind their general tendency to delay investments when exogenous performance improvements can be anticipated~\parencite{rubin_review_2015}. In essence, if a model with enough foresight identifies that cost decrements or efficiency improvements occur due to external factors, it may delay deployment until said improvements materialise without representing the mechanisms that drive technological improvement in reality~\parencite{mannhardt_understanding_2024, heuberger_power_2017}. We can call this \textit{opportunistic investment behaviour}.

As a way to counteract opportunistic behaviour, modelling exercises often incorporate constraints affecting investment trends with the aim to guide solutions toward what the modellers see as preferable or more realistic. Examples include, first, constraints on deployment growth~\parencite{keppo_short_2010}; in other words, the amount of new capacity of a given technology that is considered realistic to deploy between two investment milestones is arbitrarily fixed. Second, knowledge stock effects~\parencite{mannhardt_understanding_2024}, which consists of non-monetary expertise a given nation acquires or loses in relation to a technology, enabling a model to endogenously alter deployment growth based on recent trends. Third, endogenous learning~\parencite{messner_endogenized_1997}, namely a mathematical relation between the amount of capacity deployment of a given technology and its cost, which reduces the more the technology becomes deployed, mimicking real-world technological learning and cost reductions~\parencite{schmidt_future_2017}. Fourth and final, endogenous early retirement~\parencite{manuel_high_2022}; where investments may be retired before the end of their technical lifetimes when doing so is financially attractive, allowing newer technologies to diffuse earlier. Their use varies across modelling frameworks and studies. For example, TIMES offers both endogenous learning and early retirement as optional features that depend on user configuration~\parencite{loulou_documentation_2016}; similarly, ESO-X, MESSAGE, and PyPSA have been employed both with and without endogenous learning as a feature~\parencite{heuberger_power_2017, messner_endogenized_1997, zeyen_endogenous_2023}. 

\section{Methods}
\label{sec:methods}

% Scaffold:
% - Introduction to the protocol used and similar reviews using quantitative approaches.
% - Section detailing the study gathering process and elimination. Make sure to justify the use of an active learning tool and explain its human-in-the-loop nature. 
% - Section detailing the data collection stage, which is mostly the questionnarie.
% - A section discussing model analysis?

Building and expanding on the key methodological concepts outlined in~\Cref{sec:theory}, we conduct a systematic review of energy systems modelling literature following the \gls{prisma} protocol \parencite{page_prisma_2021}, with the aim of quantitatively assessing how pathway modelling studies published across multiple decades approach such methodological dilemmas. Our protocol-based analysis also offers a structured and replicable approach to reviewing this field of research, which may be easily and consistently updated by others in the future. We build upon other reviews of energy systems modelling literature that take quantitative approaches~\parencite{lopion_review_2018, behrens_reviewing_2024, chang_trends_2021} by narrowing the subject to a particular aspect of \glspl{esom} (i.e., pathway exercises) while ensuring our inclusion criteria captures a broad range of studies. The following sections detail our procedure: first, how our sample of studies was identified and filtered, and second, the data collection processes.

\subsection{Gathering the sample of studies}

Data gathering started by developing an appropriate search string to capture terminology often used to describe pathway studies using \glspl{esom}. As seen in \Cref{tab:search-terms}, we combined queries related to four particular aspects of interest: (i) the systems under study, which was energy or power systems; (ii) the models used, namely power and energy planning models; (iii) the modelling approach, which was optimisation via mathematical programming; (iv) the type of exercise, in this case pathways or multi-stage investments. When relevant, a proximity operator---\textit{PRE/n}---was used to capture cases where terminology might vary slightly due to sectoral scope or other particularities (e.g., \textit{energy PRE/2 model*} will identify studies using \textit{energy systems model}, \textit{energy modelling} and \textit{energy-economy-environment models} as valid). We purposefully avoid restrictions on the publication date of papers, subject area, or journal ranking in order to diminish biases.

\begin{table}[bt]
\centering
\caption{Terms used to query article titles, abstracts and keywords in the Scopus database~\parencite{elsevier_bv_scopus_2025}.}
\label{tab:search-terms}
\resizebox{\columnwidth}{!}{%
\begin{tabular}{lll}
\hline
Aspect   & Operator & Query                                             \\ \midrule
System   & -        & ( energy OR power OR electric* ) PRE/2 system     \\ 
Model    & AND      & ( energy OR power OR electri* OR planning )       \\
         &          & PRE/2 model*                                      \\
Approach & AND      & optim* OR "bottom up" OR "cost competitive*"      \\
Exercise & AND      & pathway OR "multi* period" OR "multi* stage"      \\
         &          & OR "long term" OR scenario OR "transition path"   \\
         &          & OR "generation expansion" OR "capacity expansion" \\ \bottomrule
\end{tabular}%
}
\end{table}

\begin{figure}[bt]
    \centering
    \includegraphics[width=1\linewidth]{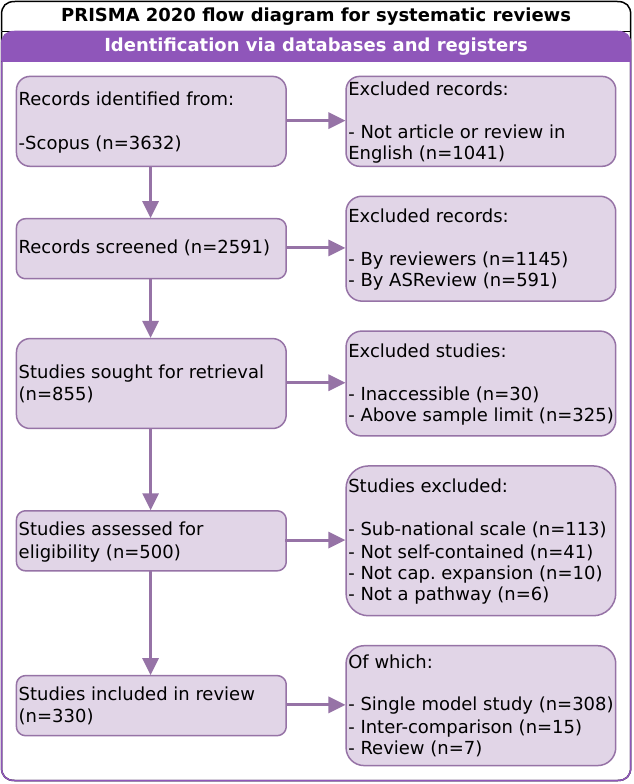}
    \caption{\gls{prisma} diagram depicting each stage in the systematic review, including the number of studies and causes for elimination. Adapted from \textcite{page_prisma_2021}.}
    \label{fig:prisma}
\end{figure}

\Cref{fig:prisma} depicts each stage in our study identification procedure. The Scopus database~\parencite{elsevier_bv_scopus_2025} was used as the primary source of data, resulting in a total of 3632 records after being consulted on the 6\textsuperscript{th} of January 2025. This sample was reduced to 2591 records by filtering out publications that were not written in English or that were not scientific articles or reviews. Although pathways are often featured in grey literature and government reports, both potentially valuable sources of information, the focus of this review is on assessing practices and methods in peer-reviewed scientific literature, meaning that such documents were outside the scope of this study.

To aid in the screening of this large sample of studies, we used ASReview LAB version 1.6.6~\parencite{developers_asreview_2025}, a machine-learning tool that uses active learning to assist systematic reviews and meta-analyses in an efficient and transparent manner~\parencite{van_de_schoot_open_2021}. The tool has a human-in-the-loop approach, meaning reviewers are ultimately the ones guiding record selection by reading article titles and abstracts; the tool re-arranges the order in which record titles and abstracts are presented to the reviewer to speed up the review process. During this stage, we selected studies mentioning the use of \glspl{esom} in the context of pathways or capacity expansion. If the title and abstract of a study omitted such information, the main body of the text was briefly consulted to assess the study's relevance. To diminish the risk of the ASReview presets biasing our selection, we split the screening into two phases using different classifiers and feature extractors, and an additional validation phase using randomised studies (see~\Cref{fig:screening}). Our screening yielded a total of 855 records, which we further reduced to 500 by removing 30 records that were inaccessible and selecting the 500 studies with the highest ASReview priority score from among the remaining 825.

\subsection{Assessment and data collection}

We developed a coded questionnaire, depicted in \Cref{tab:questions}, to extract condensed data from each study. Coders were required to mark the sentence or paragraph that made them give a specific answer. Inter-coder reliability was assessed and is reported in the results. We subdivided the questionnaire into three stages, with studies that did not meet certain criteria being excluded from further analysis: (i) an eligibility stage aimed to improve the homogeneity of our sample; (ii) a classification stage aimed at separating single-model studies from reviews and model inter-comparisons; and (iii) an analysis stage which extracted technical information from single-model studies.

The eligibility stage and the classification stage were utilised to improve the comparability of the studies in our sample. To do so, we required studies to have a national or super-national scope, removing models depicting subnational entities like states~\parencite{wei_deep_2013} or cities~\parencite{liu_importing_2022}. We similarly removed studies that either did not use an \gls{esom} or cases where its influence was difficult to establish in a self-contained way. This meant removing bi-directionally coupled models solved in iterations~\parencite{strachan_hybrid_2008} and multi-model setups~\parencite{li_decoupling_2024}, since establishing cause-and-effect in these setups is difficult and complicates comparisons. Models whose aim was tangential to energy planning, such as car stock models~\parencite{mulholland_technology_2017}, were also removed. Finally, we ensured that studies that did not feature pathways were withdrawn (e.g., studies optimising one or several ``snapshot'' years~\parencite{berntsen_ensuring_2017}. The remaining sample of studies was classified into three categories: studies utilising a single \gls{esom}, literature reviews, and model inter-comparisons.

Our analysis stage focused on extracting data from our sample of single \gls{esom} studies. First, we classified studies qualitatively on multiple details, including the modelling framework used, and the type of formulation employed (e.g., \gls{lp}, \gls{milp}, etc.). Second, we extracted spatial details, including the region under study (see \Cref{tab:world-regions}) and the model's spatial resolution ($r \in R$). Third, the temporal details of each study were identified, including: the foresight utilised, the horizon of the exercise from baseline to target year---noting modelling extensions beyond said target year if stated---, the total number of milestones modelled ($|M|$), and the total number of timesteps ($|T|$). As a general rule, we assumed that the baseline year was used only as reference and was not included in the optimisations (e.g., if a study stated ``modelled from 2015 to 2050 in 5-year steps'', then $|M| = (2050-2015)/5 = 7$). Fourth, and last, we made note of the use of endogenous learning.

% \subsection{Data evaluation}

% Uncertainty was obtained via a cluster bootstrap by year, which resamples years as blocks to preserve within-year dependence and prevent dense recent years from dominating the confidence intervals.

\section{Results}
\label{sec:results}

\subsection{Descriptive analysis}
\label{subsec:results:descriptive}

\begin{figure*}[bt]
  % Top row (50% + 50%)
  \begin{minipage}[t]{.5\linewidth}
    \centering
    \includegraphics[width=\linewidth]{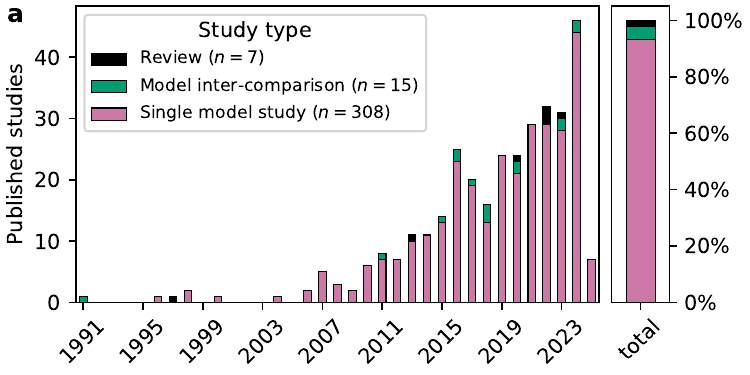}
    \phantomsubcaption\label{fig:study-type}
  \end{minipage}
  \begin{minipage}[t]{.5\linewidth}
    \centering
    \includegraphics[width=\linewidth]{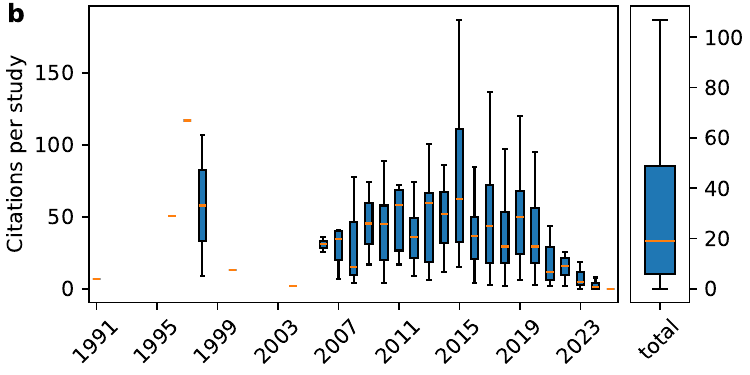}
    \phantomsubcaption\label{fig:study-citations}
  \end{minipage}
  % Bottom row (50% + 50%)
  \begin{minipage}[t]{1\linewidth}
    \centering
    \includegraphics[width=\linewidth]{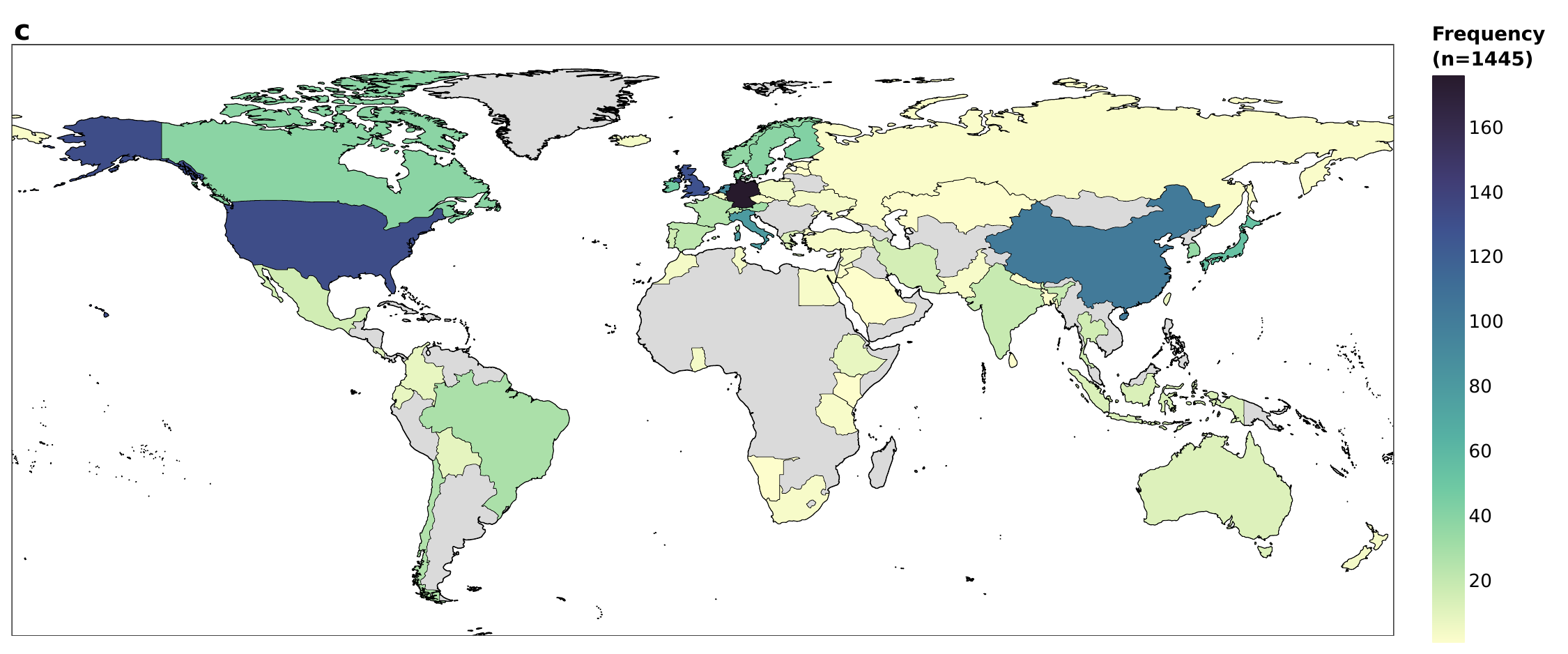}
    \phantomsubcaption\label{fig:study-authorship}
  \end{minipage}
  \caption{Bibliographic trends in our full sample of studies ($n=330$). \textbf{a} Study type  by year of publication (\textbf{Q5}). \textbf{b} Citations per study by year of publication. Boxes show the median and interquartile range (25–75\%), whiskers extend to 1.5 of the interquartile range, and outliers are omitted. \textbf{c} Author affiliation by country, counting each co-author per study separately. Countries in grey had no authorship in our sample.}
  \label{fig:bibliometric-summary}
\end{figure*}

Our systematic review identified 330 publications: 308 single-\gls{esom} studies, 7 review articles, and 15 model inter-comparisons. The publication years range from 1991 to 2025, with a persistent upward trend in publications through time (\Cref{fig:study-type}). Studies published before 2000 are scarce, possibly due to terminology not captured by our search strategy. After 2000, the number of studies increases from 2 in 2006 to 46 in 2024, reflecting the growing scientific and policy interest in energy system decarbonisation. Our analysis showed a relatively steady mean citation count of 50.3 for studies between 2006-2020, with a wide spread of $\pm 50.1$ SD (\Cref{fig:study-citations}). Citations decline from 2021 onwards, which is to be expected due to recency. It should be noted that the search was conducted in January 2025, explaining the low number of studies in that year ($n=7$).

Studies in our sample are distributed across 93 journal sources, mostly focused on disciplines such as energy research, policy, power systems, and sustainability. Among these, thirteen journals account for 67\% of the sample with \textit{Applied Energy} ($n=40$) and \textit{Energy} ($n=35$) holding the largest shares and relatively steady publication history (\Cref{fig:biblio-journal}). Conversely, \textit{Energy Policy} ($n=30$) was the third largest and showed a decreasing trend of publications within our sample in the latest half-decade. A total of 59 journals held a single article within our sample, and were composed mostly of journals of specialised technical topics (e.g., geothermal engineering, solar photovoltaics, buildings, etc.) or a dedicated regional scope (e.g., Türkiye, Canada, South Africa). Neither of these is surprising given the wide breadth of technologies included in \glspl{esom} and our eligibility criteria.

We similarly assessed the distribution of author affiliation within our sample, which showed that a reduced number of countries hold significant influence over this field of literature (\Cref{fig:study-authorship}). Of the total of 1445 identified co-author counts, ten nations accounted for 62\% in terms of author affiliation, with Germany (12.2\%), the U.S. (9.1\%), the U.K. (8.9\%), China (7.1\%) and the Netherlands (6.3\%) being the largest among them. Co-authors with European affiliations are over-represented in studies, with a total share of 58.1\% of the total ($n=839$), followed by Asia with 19.6\% ($n=283$), and North America with 13.6\% ($n=197$). In contrast, Oceania, Africa and South America are under-represented with 1.1\%, 2.1\%, and 5.5\% respectively (see \Cref{tab:co-authorship-regions,tab:co-authorship-countries} for details).

\subsection{General qualitative features of ESOMs}
\label{subsec:results:features}

\begin{figure*}[bt]
    \centering
    % LEFT COLUMN
    \begin{minipage}[b]{.5\linewidth}
        \centering
        \includegraphics[width=\linewidth]{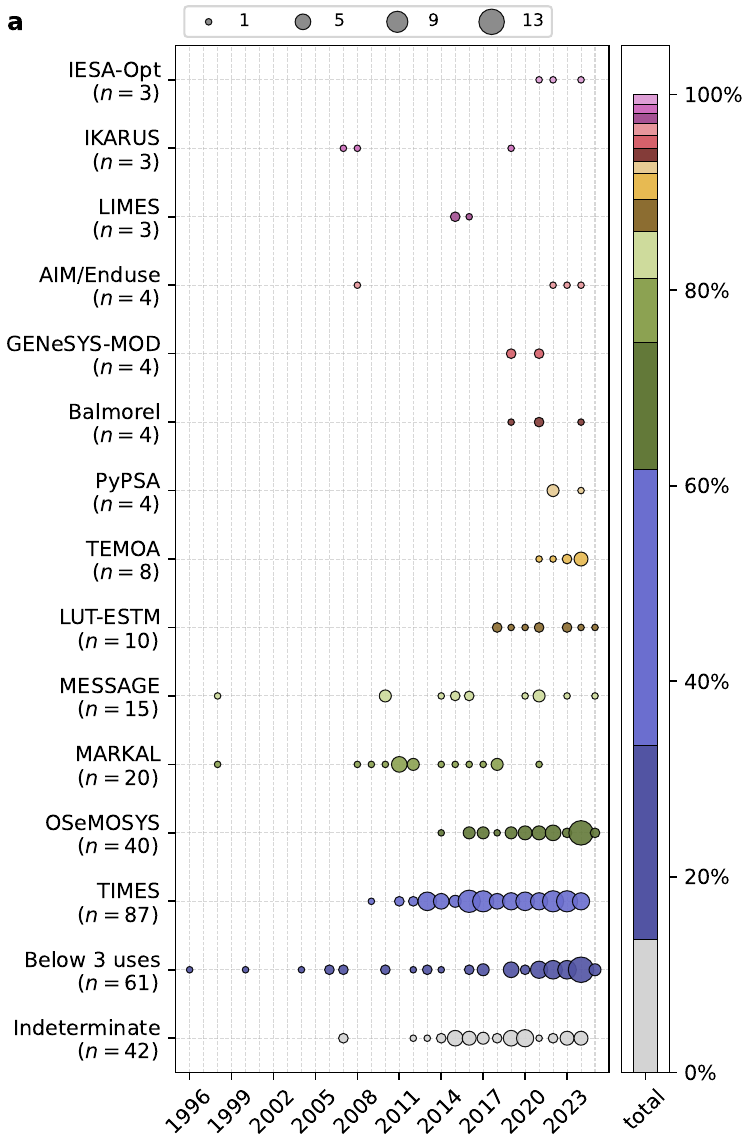}
        \phantomsubcaption\label{fig:model-framework}
    \end{minipage}%
    \begin{minipage}[b]{.5\linewidth}
        \centering
        \includegraphics[width=\linewidth]{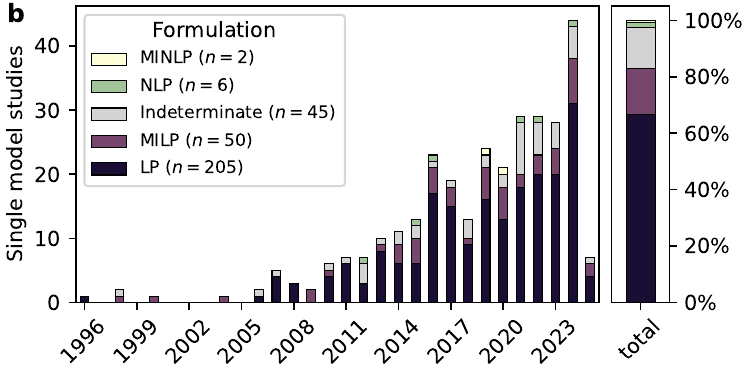}
        \phantomsubcaption\label{fig:model-formulation}
        \includegraphics[width=\linewidth]{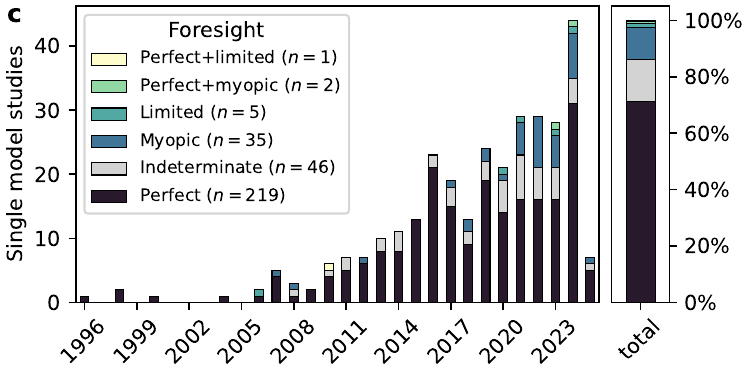}
        \phantomsubcaption\label{fig:model-foresight}
        \includegraphics[width=\linewidth]{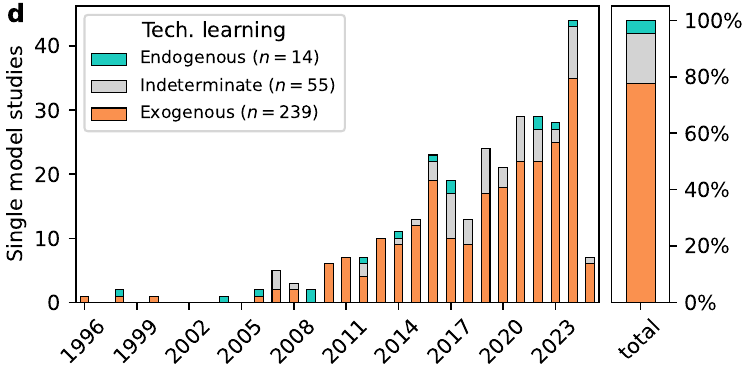}
        \phantomsubcaption\label{fig:model-learning}
        
    \end{minipage}
    \caption{Summary of model characteristics in single-model studies by year of publication ($n=308$). \textbf{a} Model framework used. Frameworks with lower usage ($n < 3$) have been aggregated into the ``Below 3 uses'' category. \textbf{b} Model formulation as identified by coders. \textbf{c} Foresight approach as identified by coders. Cases where authors contrasted multiple foresight formulations have been given their own categories. \textbf{d} Technological learning approach as identified by coders.}
    \label{fig:model-features}
\end{figure*}

We evaluate single-model studies ($n=308$) on multiple qualitative aspects including model framework and formulation, foresight, technology learning approach and modeller region. This section reports substantive characteristics only; reporting completeness (i.e., indeterminate items) is analysed in \cref{subsec:results:transparency}.

\paragraph{Modelling frameworks}
We identified 63 named modelling frameworks, 14 of which were used more than three times (\Cref{fig:model-framework}). TIMES was the most used framework at 28\% ($n=87$), which expands to 34\% when combined with its predecessor MARKAL ($n=20$)~\parencite{Loulou.R.2008_ETSAPTIAMTIMESIntegrateda}. OSeMOSYS was the second most used framework at 13\% ($n=40$), which increases to 14\% when accounting for GENeSYS-MOD ($n=4$), which was based on the same code base~\parencite{bartholdsen_pathways_2019}. Other frameworks included MESSAGE ($n=15$), LUT-ESTM ($n=10$) and TEMOA ($n=8$). Notably, the number of frameworks below 3 uses increased sharply in the last half-decade (20\%, $n=20$), indicating growing diversification in the field. 

\paragraph{Mathematical formulation}
This category displayed a broadly consistent preference towards tractable formulations. Linear programming (\gls{lp}) dominated (67\%, $n=205$), followed by \gls{milp} (16\%, $n=50$). Non-linear approaches were rare (2.5\%, $n=8$), which is to be expected as it implies higher computation costs~\parencite{kotzur_modelers_2021}. There is no strong evidence of complex formulations becoming more popular, with \gls{lp} remaining the preferred approach across the years.

\paragraph{Foresight}
Perfect foresight was the most used approach (71\%, $n=219$), with myopic rolling horizon being a far second (11\%, $n=35$), and limited-foresight exercises a scarce third (2\%, $n=5$). Myopic exercises have grown more common in the past half decade, with 69\% ($n=22$) of such studies appearing between 2020 and 2025. Only a few studies compared foresight approaches directly ($n=3$)~\parencite{limpens_energyscope_2024, keppo_short_2010, marchenko_development_2023}. Overall, foresight choice is quite binary between perfect and myopic foresight, with limited approaches remaining rare occurrences.

\paragraph{Endogenous learning}
Endogenous learning was uncommon (5\%, $n=14$), implying that most models have a weakened causal link between technology deployment and cost reductions from economies of scale. This type of setup is known to lead to delayed system change and a potential for the model to exhibit opportunistic behaviour~\parencite{rubin_review_2015}. The trend is unsurprising, as endogenous learning increases computational complexity, either by introducing non-linear relations or necessitating iterative model runs. Despite these challenges, our sample included cases where modellers employed more involved approaches---such as Benders decomposition---to enable endogenous learning in models with high operational resolution~\parencite{felling_multi-horizon_2022}. Such combinations of decomposition techniques and endogenous learning have been highlighted by other studies as promising avenues for improvement~\parencite{behrens_reviewing_2024}.

\begin{figure}
    \centering
    \includegraphics[width=1\linewidth]{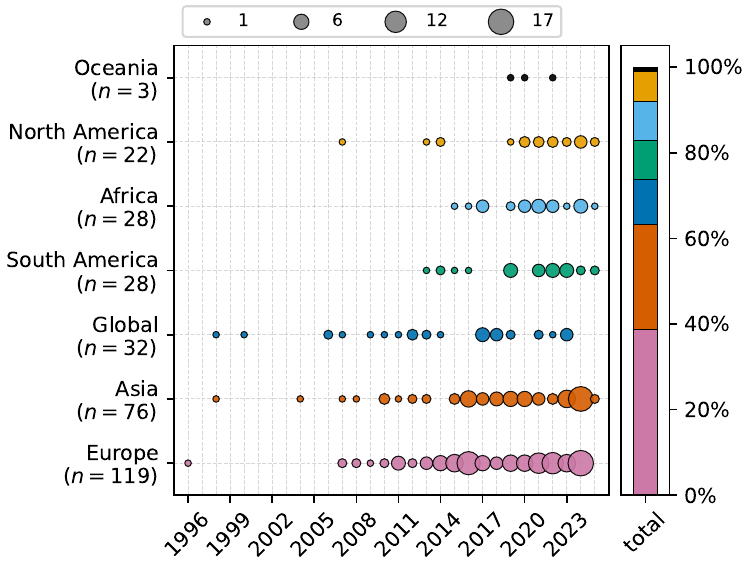}
    \caption{World region modelled in single-model studies by year of publication ($n=308$). See \Cref{tab:world-regions} for further details on country assignation.}
    \label{fig:model-region}
\end{figure}

\paragraph{Modelled region}
\Cref{fig:model-region} shows how the regional focus of these studies generally reflects the authorship trends identified in \cref{subsec:results:descriptive}, with Europe (39\%, $n=119$) and Asia (25\%, $n=76$) being the main areas of focus, followed by global models as the third largest category (10\%, $n=32$). North America (7\%, $n=22$) had a lower share than its co-authorship counterpart, which may be explained as an artefact of our decision to exclude sub-national studies, which therefore excludes studies of sub-national U.S. grids such as ERCOT~\parencite{us_epa_us_2022}. Of the remaining regions, South America (9\%, $n=28$), and Africa (9\%, $n=28$) had larger shares than their co-authorship, and Oceania had a roughly equal amount (1\%, $n=3$).
% No study in our sample had an indeterminate geographic area of interest, since such cases were removed during the eligibility stage.

\subsection{Degree of transparency}
\label{subsec:results:transparency}

Our questionnaire assessed reporting clarity for 12 items, allowing coders to mark an item as \emph{indeterminate} when a study did not facilitate an unambiguous entry. \Cref{tab:transparency} summarises the indeterminate shares by item.

\begin{table}[bt]
\caption{Summary of indeterminate cases per question for single model exercises ($n=308$).}
\label{tab:transparency}
\resizebox{\columnwidth}{!}{%
\begin{tabular}{@{}lllr@{}}
\toprule
\textbf{Subject}    & \textbf{Question}    & \textbf{Total (\%)} \\ \midrule
Model specification & Modelling framework  & 42 (13.6\%)         \\
                    & Model name           & 163 (52.9\%)        \\
                    & Formulation          & 45 (14.6\%)         \\ \midrule
Spatial detail      & World region         & 0 (0.0\%)           \\
                    & Number of regions    & 46 (14.9\%)         \\ \midrule
Temporal detail     & Foresight            & 46 (14.9\%)         \\
                    & Baseline year        & 8 (2.6\%)           \\
                    & Target year          & 6 (1.9\%)           \\
                    & Horizon year         & 8 (2.6\%)           \\
                    & Number of milestones & 12 (3.9\%)          \\
                    & Number of timesteps  & 85 (27.6\%)         \\ \midrule
Investment dynamics & Learning approach    & 55 (17.9\%)         \\ \bottomrule
\end{tabular}%
}
\end{table}

Studies did not assign a name to the framework utilised in 14\% of cases ($n=42$). This is not necessarily problematic as models can be tailor-made for a specific study (making naming irrelevant), authors may prefer not to assign a name, or journal policies may discourage naming altogether. However, it implies that at least 86\% of studies used or developed a framework intended for reuse, assuming that naming signals that intention.

Geographic scope was always identifiable---by design, as ambiguous regions were excluded during screening. Items related to a model's horizon (baseline year, target year, horizon year, number of milestones) were coded in most cases, each with <4\% indeterminate.

Items central to determining computational tractability showed more opacity. Mathematical formulation (e.g., \gls{lp}, \gls{milp}) and foresight were indeterminate in 15\% of studies, which is compounded by additional opaqueness regarding the number of regions modelled (15\%) and the number of operational timesteps (27.2\%). These omissions hinder the ability of readers to interpret both operational detail and the computational burden of models.

Endogenous learning was indeterminate in 18\% of studies, reflecting difficulties in confirming its absence. Despite previous reviews highlighting its importance~\parencite{decarolis_formalizing_2017}, it is not a widely used method, likely due to computational burden~\parencite{behrens_reviewing_2024}. Most studies implicitly assumed exogenous learning, increasing the burden of proof of systematic exercises like this one.

The numbers in \Cref{tab:transparency} also mask other difficulties that were hard to quantify. Studies frequently left key model aspects open to interpretation, and it was not uncommon for authors to cite a framework without specifying configuration or resolution, even though most frameworks allow multiple mathematical formulations, temporal/spatial resolutions, and optional constraints. In other cases, readers were referred to multiple model versions with differing configurations, or to grey literature that was effectively inaccessible due to link rot or access restrictions. Inter-coder analyses showed a similar pattern: model formulation, foresight and learning approach produced disagreement, which completely disappeared when indeterminate cases were excluded (\Cref{tab:intercoder-qualitative}). That is, coders disagreed about clarity rather than the feature itself when it was disclosed. Quantitative questions showed comparable issues, with the number of model milestones proving particularly challenging and often leading coders to report different counts when the long-term resolution was vaguely stated.

\subsection{Quantitative analysis of model resolution}
\label{subsec:results:resolution}

% This section compresses our sample to studies that were transparent in terms of resolution and horizon (180 for resolution, I have yet to do the horizon analysis). The goal is to compare these models are in an equivalent way, and then identify trends in the field might be affected by problems we identified.

% \begin{itemize}
%     \item How do different foresight approaches deal with resolution? Are models of each type improving their resolution?
%     \item Is there an increasing risk that models might be affected by end-of-horizon effects (e.g., models with short horizons without extension)?
%     \item Are there many cases where year aggregation might affect results (i.e., inter-milestone distortions)?
% \end{itemize}

\begin{figure*}[!hbt]
    \centering
    % Row 1 — timesteps (L), regions (R)
    \begin{subfigure}[b]{.49\linewidth}
        \centering
        \includegraphics[width=\linewidth,trim=0 0 0 0,clip]{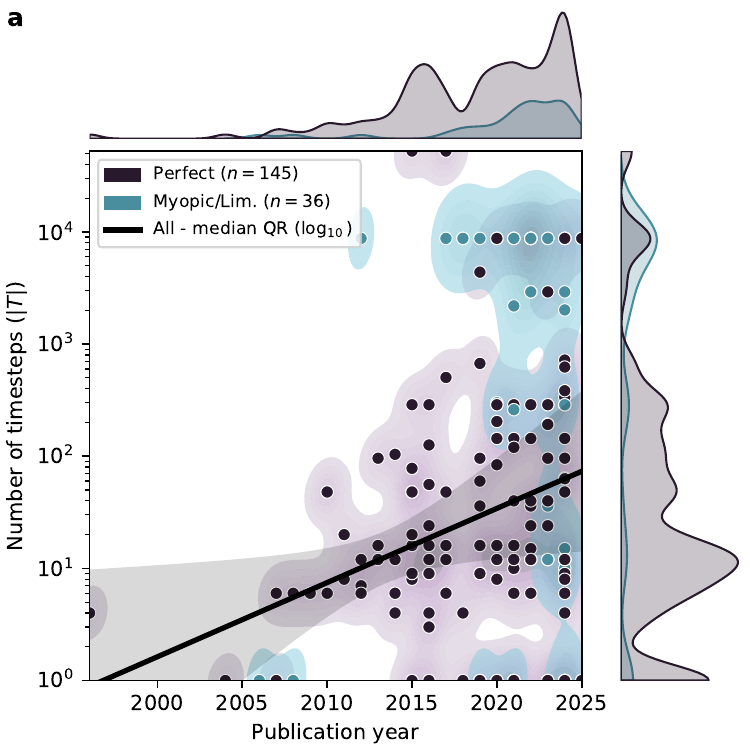}
        \phantomsubcaption\label{fig:resolution-timesteps}
    \end{subfigure}%
    \begin{subfigure}[b]{.49\linewidth}
        \centering
        \includegraphics[width=\linewidth,trim=0 0 0 0,clip]{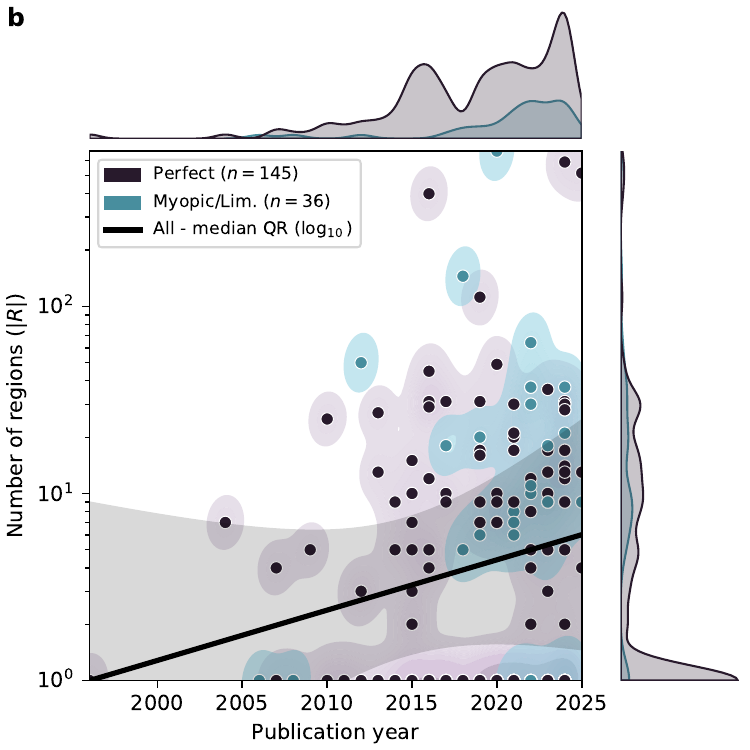}
        \phantomsubcaption\label{fig:resolution-regions}
    \end{subfigure}
    
    \vspace{-0.9\baselineskip}
    % Row 2 — milestones (L), combined (R)
    \begin{subfigure}[b]{.49\linewidth}
        \centering
        \includegraphics[width=\linewidth,trim=0 0 0 0,clip]{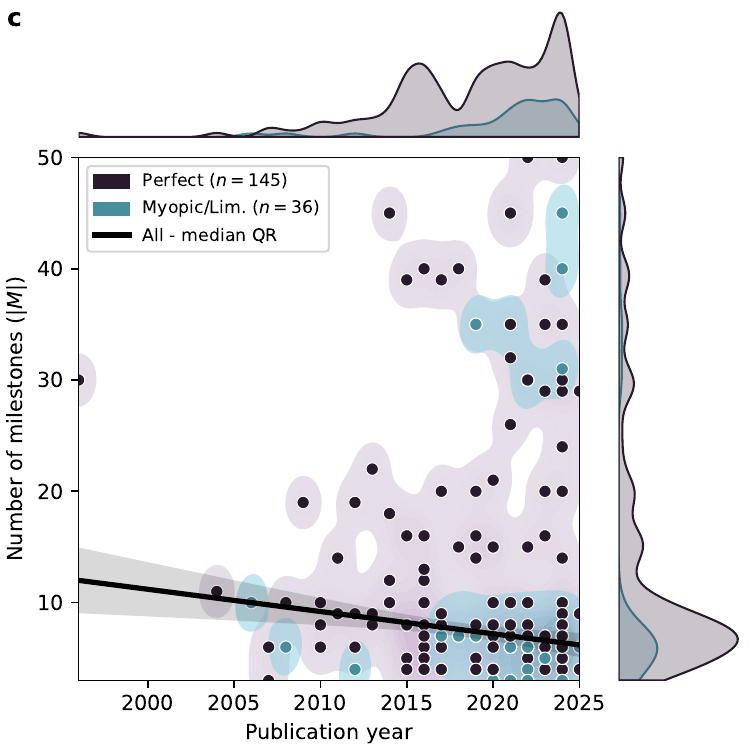}
        \phantomsubcaption\label{fig:resolution-milestones}
    \end{subfigure}%
    \begin{subfigure}[b]{.49\linewidth}
        \centering
        \includegraphics[width=\linewidth,trim=0 0 0 0,clip]{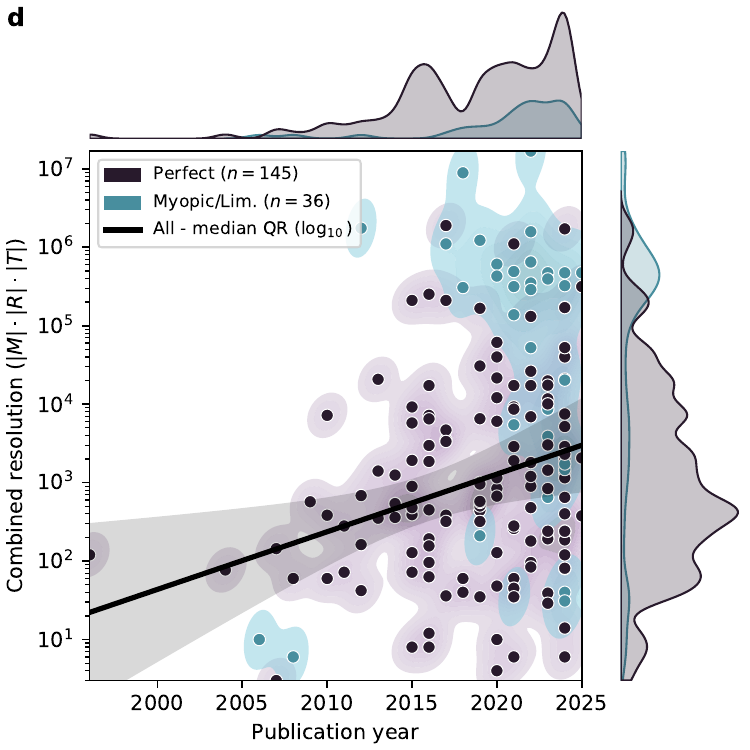}
        \phantomsubcaption\label{fig:resolution-combined}
    \end{subfigure}
    
    \caption{Dimensional characteristics of single-model studies with sufficient transparency by publication year and foresight approach ($n=181$). Solid lines are median quantile regressions ($\tau = 0.5$) with 95\% confidence bands from a pairs bootstrap. Sides panels show kernel density estimates by foresight category. \textbf{a} Operational resolution (total timesteps $|T|$). \textbf{b} Spatial resolution (total regions $|R|$). \textbf{c} Long-term resolution (total milestones $|M|$). \textbf{d} Evolution of combined resolution ($|M|\cdot |R| \cdot |T|$)---an indicator of study setup, not computational burden. Y-axes use $log_{10}$ scaling in a, b, and d; linear scale in c.}
    \label{fig:resolution-by-pubyear}
\end{figure*}

Our large sample of studies allows a unique opportunity to assess the evolution of multiple aspects of \glspl{esom} across the years. To do so, we isolate a subset of studies that provided sufficient data for the following items: model foresight, number of milestones ($|M|$), number of regions ($|R|$), number of operational timesteps ($|T|$), baseline year, target year, and horizon year. This resulted in a reduced subset of 181 studies (59\% of 308). We group studies into those with perfect foresight ($n=145$) and those using myopic or limited foresight ($n=36$).

We evaluate trends in the dimensionality of \glspl{esom} in our subset in recent decades (\Cref{fig:resolution-by-pubyear}), revealing distinct trends despite the wide heterogeneity of these models. In particular, there is an increasing preference towards higher short-term resolution---in both temporal and spatial aspects---and a decreasing trend in long-term resolution.

Short-term temporal resolution---measured as the number of operational timesteps $|T|$---shows clear increases over time (\Cref{fig:resolution-timesteps}), with an annual multiplicative change of 16.4\% (95\% CI [7.2\%, 31.8\%]; $n=181$), equivalent to doubling roughly every 4.6 years (95\% CI [2.5, 10]). Despite this upward trend, dispersion remains large, indicating substantial heterogeneity across modelling studies. Foresight strongly influences operational resolution: studies with myopic or limited foresight tend to favour near-hourly annual operation at a median of 2920 (IQR 31-8760; $n=36$), with studies using myopic versions of PyPSA and LUT-ESTM often favouring full hourly resolution (\Cref{fig:resolution-by-framework}). Conversely, perfect foresight models are more dispersed in their approach and use lower resolutions at a median of 16 timesteps (IQR 6-120; $n=145$). This variability also holds within frameworks: models built with OSeMOSYS, TEMOA and TIMES vary by over two orders of magnitude in this metric (\Cref{fig:resolution-by-framework}).

Spatial resolution---total regions $|R|$---shows a modest upward tendency over time. 
The median trend indicates an annual increase of 6.4\% (95\% CI $-3.3$-14.7\%; $n=181$). Over a decade, this compounds to about 86\% higher on the median. Because the confidence interval includes 0\%, the evidence is suggestive rather than definitive (\Cref{fig:resolution-regions}). There is a clear trade-off between foresight and spatial detail: models either pursue finer spatial resolution at the cost of less or no anticipation in investments, or retain anticipation by neglecting energy transmission costs and weather differences across regions (\Cref{fig:resolution-aggregate}). By foresight type, perfect foresight studies tended to use single-region ``copperplate'' setups (median 1, IQR 1-10; $n=145$), a pattern that persists even in recent work, whereas myopic and limited foresight models were more spatially disaggregated (median 9, IQR 5.75-23.24; $n=36$).

\Cref{fig:resolution-milestones} shows that long-term temporal resolution---milestones $|M|$ across the investment horizon---declines over time. The median trend is -0.2 milestones per year (95\% CI [-0.08, -0.27]; $n=181$). This corresponds to roughly two fewer milestones per decade assuming milestones are spaced five years apart ($\mathbf{ML}=5$). Aggregate metrics suggest that this is the only dimension in which models with perfect foresight utilise higher resolutions (median 8 milestones, IQR 6-16) than models with myopic or limited foresight (median 6 milestones, IQR 5.75-7), although in general this aspect remains coarse in both approaches (\Cref{fig:resolution-aggregate}).

To answer the question of whether or not there is a focus on increasing resolution in these studies as a whole, we look at their combined resolution (the product $|M|\cdot |R| \cdot |T|$) in \Cref{fig:resolution-combined}. Importantly, this measure reflects study setup rather than the computational burden of a model, since myopic and limited-foresight models do not resolve all milestones simultaneously. The increases in short-term resolution ($|T|$) dominate, leading to a similar annual increase of 18.4\% (95\% CI [10.53\%, 34.59\%], which would compound to about a 443\% increase on the median over a decade. Dispersion is large (log10-MAE 1.13) due to the substantial heterogeneity across studies and years. This large increase must be put into context, as this category has the most clear outlines between foresight categories (\Cref{fig:resolution-aggregate}). Perfect foresight studies had an aggregate median of $10^{2.81}$ (IQR [$10^{2.16}$, $10^{3.81}$]) which was about three orders of magnitude below that of myopic or limited foresight models ($10^{5.49}$, IQR [$10^{3.52}$, $10^{5.72}$]). Looking at specific frameworks (\Cref{fig:resolution-by-framework}) reveals a few outliers, with the IESA-Opt model~\parencite{beres_impact_2024} and the model of \textcite{komiyama_optimal_2017} as standout perfect foresight cases with a higher combined resolution than most myopic models solved at an hourly operational resolution.

\subsection{The heuristics of optimised pathways}

\begin{figure*}
    \centering
    \begin{subfigure}[b]{.49\linewidth}
        \centering
        \includegraphics[width=\linewidth]{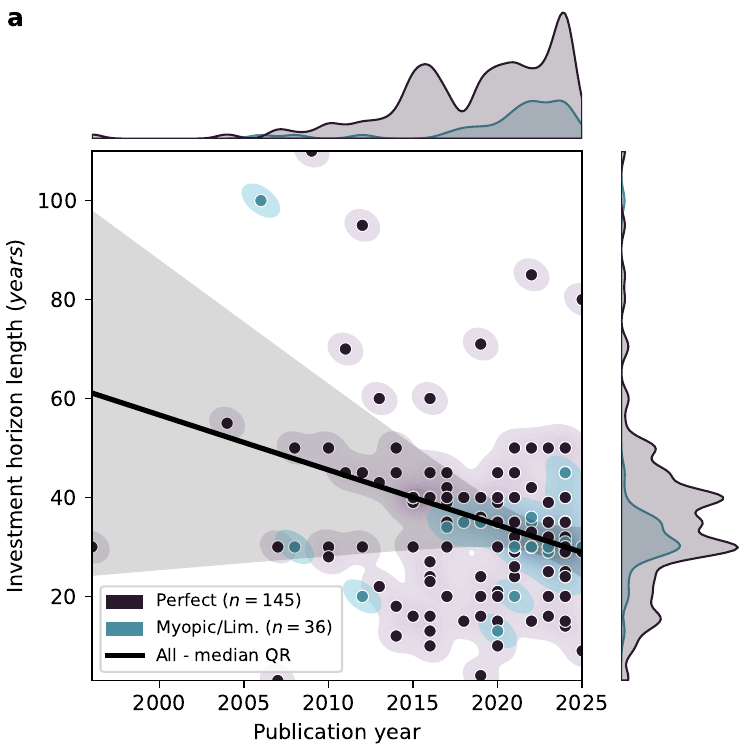}
        \phantomsubcaption\label{fig:resolution-horizon-length}
    \end{subfigure}
    \begin{subfigure}[b]{.49\linewidth}
        \centering
        \includegraphics[width=\linewidth]{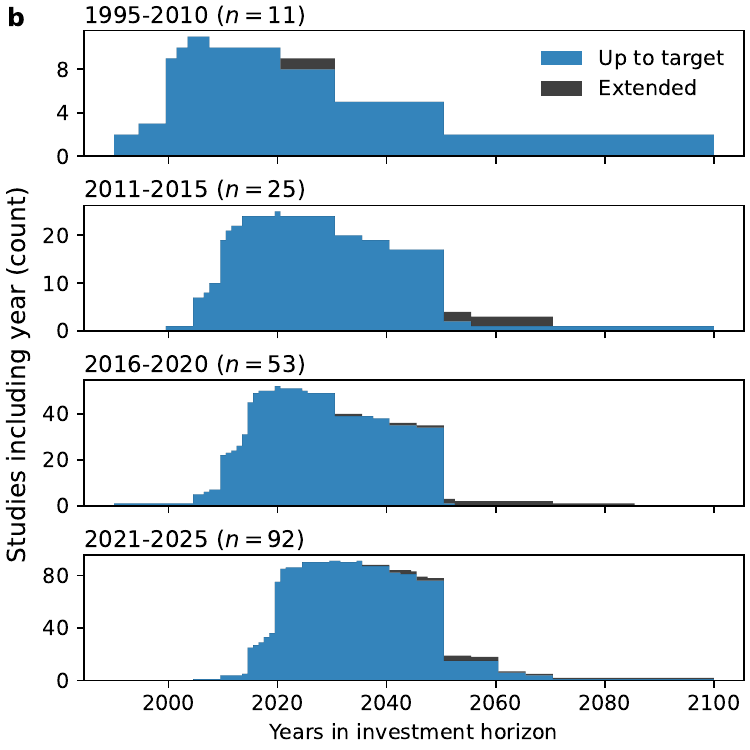}
        \phantomsubcaption\label{fig:resolution-horizon-extension}
    \end{subfigure}
    \caption{Evaluation of long-term characteristics of single-model studies with sufficient transparency that relate to end effects ($n=181$). \textbf{a} Length of the modelled investment horizon by year of publication and foresight approach. Solid line is a median quantile regression ($\tau = 0.5$) with 95\% confidence bands from a pairs bootstrap. \textbf{b} Counts of each modelled year, subdivided into groups of close publication date. Cases where authors extended the horizon beyond the target year of interest have been highlighted in black.}
    \label{fig:resolution-investment-horizon}
\end{figure*}

\Cref{sec:theory} described two types of issues that can surface in long-term exercises: (i) end effects, which occur when models with foresight do not extend beyond the year of interest, and (ii) inter-milestone effects, which occur when years in the investment horizon are aggregated into groups (i.e., $\mathbf{ML}_m > 1$). Although the presence of said issues broadly depends on the experimental setup of each study and the formulation of each model, we can assess modeller practices that surround them. 

As noted in \cref{subsec:end-effects}, end effects are mitigated---but not eliminated---by heuristics such as salvage values or heavy discounting; robust handling typically also requires extending the model beyond the target year \parencite{grinold_model_1983}. To assess practice, we examine two indicators: (i) the length of the investment horizon (total years spanned; \Cref{fig:theory-horizons}), and (ii) whether authors explicitly state an extension beyond the target year.

For the first indicator, \Cref{fig:resolution-horizon-length} suggests a decrease in horizon length over publication years: a linear median regression estimates -1.11 years per calendar year (95\% CI [-1.25, 0.00]). For the second, we grouped studies with close publication proximity and counted explicit horizon extensions (\Cref{fig:resolution-horizon-extension}). Only 15 studies reported an extension (8.2\% of $n=181$), including frameworks with distinct approaches for end effect mitigation. These include models with both salvage values (OSeMOSYS $n=6$, TIMES $n=3$), and annuity factors (SWITCH $n=1$, GridPath $n=1$); see \Cref{tab:eoh-studies} for details.

Taken together, these patterns suggest that explicit extensions are either uncommon or unreported, while planning horizons are likely shortening in the literature. There is a strong tendency to terminate analyses at years prominently featured in policy targets (e.g., 2050, 2100), with 2050 being a standout across categories. Thus, \gls{esom} studies appear increasingly short-term in scope, primarily handling end effects via heuristics (e.g., salvage values) rather than via a conscious re-structuring of the problem such as horizon extensions. These practices can bias results towards near-term solutions near important policy years, favouring technologies with short payback periods and undervaluing long-lived assets or variable renewable technologies~\parencite{krishnan_building_2017}.

\begin{figure}
    \centering
    \includegraphics[width=0.98\linewidth]{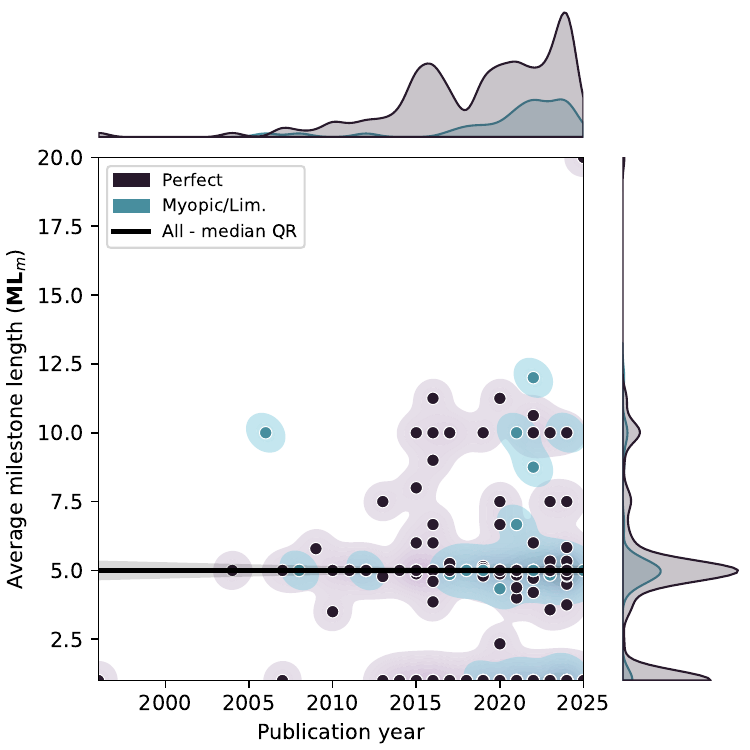}
    \caption{Long-term aggregation trends in single-model studies with sufficient transparency ($n=181$) by year of publication and foresight approach. Solid line is a median quantile regression ($\tau = 0.5$) with 95\% confidence bands from a pairs bootstrap.}
    \label{fig:year-aggregation}
\end{figure}

Inter-milestone effects depend on interactions between model structure and exogenous parameters. For example, a model with no or very coarse milestone interpolation (i.e., \Cref{fig:capacity-evolution-single,fig:capacity-evolution-square}) will be ill-suited to evaluate technologies with short lifetimes if used in combination with long milestone lengths. Thorough comparison between model formulations is beyond the scope of this review, as it requires a curated harmonisation between models to enable proper evaluation. Nevertheless, we can assess the need for such comparisons by looking at trends in model setup.

\Cref{fig:year-aggregation} suggests that time aggregation is typically chosen heuristically. Most studies either optimise every year---often intractable when other dimensions are highly resolved---or group years into multiples of five as a rule of thumb. The latter does not align with the expected decommissioning dynamics of the energy transition~\parencite{farfan_structural_2017}, and can lead to distorted results depending on the model's structure and application. As an example, the second most popular framework---OSeMOSYS---does not interpolate or represent the period between investment milestones. Its core design goal was to have a straightforward formulation~\parencite{howells_osemosys_2011}, so the choice to use discrete, non-interpolated periods is unsurprising. Nevertheless, this can lead to biased estimates of cumulative annual trends unless additional assumptions or a more detailed formulation are used to describe the evolution between milestones. If these are omitted or applied inconsistently, the resulting cumulative quantities may be systematically misestimated. This matters for common policy questions such as national carbon budgets~\parencite{van_den_berg_implications_2020}.

\section{Discussion and conclusion}
\label{sec:discussion}

We conducted a systematic review of energy system pathway exercises using \glspl{esom} in the academic literature, focusing on technical aspects of model setup that influence results, but are often left implicit. We highlight four critical aspects: i) the implications of model \textit{foresight} assumptions, ii) \textit{decision horizons} associated \textit{end effects} that can distort model outcomes, iii) tradeoffs in model resolution, and iv) investment dynamics and the need to counteract \textit{opportunistic investment behaviour} by incorporating endogenous learning. 

From this review, we reveal four broad patterns corresponding to the critical aspects identified above. For i), long-term \glspl{esom} generally fall into two categories: either perfect foresight exercises with strong anticipatory capacity but relatively low spatial and operational resolution, or, increasingly, myopic exercises that offer high spatial and operational detail but struggle to account for long-term policy goals. For ii), modellers typically follow heuristic rules of thumb when choosing the investment horizon in pathway exercises, leading to a strong trend towards shorter model horizons as key carbon-neutrality deadlines draw nearer; in many formulations, this can bias models towards short-term solutions because end effects distort investment choices. For iii), there is a clear trend towards higher operational accuracy, with finer short-term temporal resolution receiving particular attention, reflecting the need to model variable renewable technologies such as solar and wind with greater precision. For iv), the use of endogenous learning remains uncommon, likely because of computational and algorithmic complexity, leaving models with a tendency to delay investments in expectation of future price decreases, which goes against well-understood economies of scale.

This highlights several areas for improvement and further research. The choices of model foresight, combined spatial–operational resolution, and endogenous learning---points i, iii, and iv---are closely interrelated, in large part because of computational limitations, and different configurations imbue models with different biases. The current binary between perfect foresight and full myopia suggests room for improvement, as limited foresight approaches might offer a better trade-off between a model's anticipatory ability and its operational and spatial resolution, reducing the risk of lock in and stranded assets. Similarly, greater awareness is needed of potential model distortions caused by short horizons and coarse long-term aggregation. Although scarce, literature showcasing more rigorous treatment of end effects does exist~\parencite{krishnan_building_2017}, and established models like TIMES have documented the impacts of improper milestone aggregation~\parencite{lehtila_times_2016}. Model developers and users stand to benefit from adopting more systematic approaches to evaluate and contrast these aspects. Finally, there is a pressing need for greater transparency in how the long-term setup of these models is communicated: pathway exercises often leave out contextual details that are essential for interpreting their results, as has also been noted by other work on energy system modelling best practices~\parencite{decarolis_formalizing_2017, pfenninger_open_2024}.

This work has several limitations. Our search terms and article selection criteria were designed to capture and harmonise studies in order to identify modeller practices related to pathway setup, not to provide evidence about the quality of those studies per se. Other important aspects affecting the quality of \gls{esom} studies, such as operational formulations, approaches to uncertainty mitigation, and up-to-date model parametrisation, are outside the scope of this work. Furthermore, models are mutable pieces of software, meaning that the articles and model documentation we use as examples may not reflect the latest versions of those models. The critical aspects and theoretical concepts highlighted here are, however, expected to remain relevant. Finally, our review does not capture modelling setups utilised in grey literature or government reports, meaning that the assessment of pathway practices at the academic-policy interface remains an important area of future study.

Our study systematically assessed practices related to optimised energy system pathways in the academic literature, highlighting important considerations that modellers and model users should take into account when interpreting results produced by \glspl{esom}. In particular, we underscore that clear communication of the type of foresight used is essential. Myopic formulations, although useful in certain situations, should be treated with care when making statements about which system setups are efficient and effective. Similarly, careful consideration of a model's formulation is necessary when defining the decision horizon and when aggregating investment milestones, in order to diminish the impact of end effects and inter-milestone effects. Despite advancements in the operational resolution of models and a few promising highly resolved cases, the computational challenge facing \glspl{esom} remains substantial due to the need for horizon extensions, the widespread coarseness in long-term resolution, and the additional complexity induced by endogenous learning formulations.

\section*{CRediT authorship contribution statement}

\textbf{Ivan Ruiz Manuel:} Conceptualization, Methodology, Software, Validation, Formal analysis, Investigation, Data Curation, Writing - Original Draft, Visualization. \textbf{Meijun Chen:} Validation, Investigation, Data Curation, Writing - Review \& Editing. \textbf{Francesco Lombardi:} Conceptualization, Methodology, Writing - Review \& Editing, Supervision. \textbf{Stefan Pfenninger-Lee:} Conceptualization, Methodology, Writing - Review \& Editing, Supervision, Funding acquisition.

\section*{Declaration of competing interest}
The authors declare that they have no known competing financial interests or personal relationships that could have appeared to influence the work reported in this paper.

\section*{Data availability}
Curated datasets and code are publicly available on \href{https://github.com/irm-codebase/pathways_systematic_review}{Github} and in appendixes.

\section*{Acknowledgements}

This work was sponsored by the Swiss Federal Office of Energy’s ``SWEET'' call 1-2020 programme and performed in the PATHFNDR (PATHways to an Efficient Future Energy System through Flexibility aND SectoR Coupling) consortium.

\printglossary[type=\acronymtype]
\printbibliography
\end{refsection}

\clearpage
%% The Appendices part is started with the command \appendix;
%% appendix sections are then done as normal sections
\appendix
\begin{refsection}
\setcounter{figure}{0} % Reset counters
\setcounter{table}{0}
\onecolumn

\section{Discount rate dynamics}
\label{sec:discount-dynamics}

\begin{figure}[hbt]
    \centering
    \includegraphics[width=0.6\linewidth]{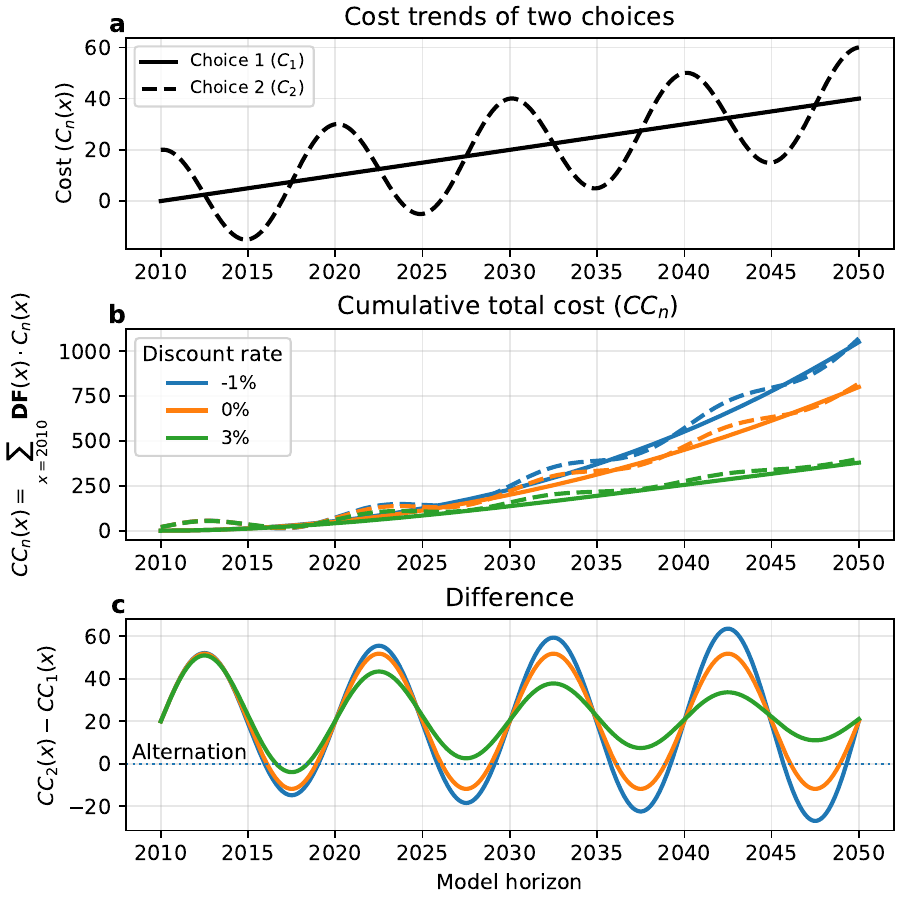}
    \caption{
    Example of discount rates influencing the length of a model's decision horizon, adapted from ~\textcite{bean_conditions_1984}. \textbf{a} Cost trends of two mutually exclusive choices. \textbf{b} Cumulative discounted cost of both choices over time. \textbf{c} Difference between cumulative costs over time.}
    \label{fig:discount-dynamics}
\end{figure}

Here we present a streamlined example of discount rates reducing decision horizons based on~\textcite{bean_conditions_1984}. There are two mutually exclusive choices: $C_1$, which steadily raises over time, and $C_2$, which follows the same steady raise but alternates around $C_1$ by a fixed amplitude (\Cref{fig:discount-dynamics}a). Because the model adds up costs only up to a chosen end year (the horizon), and because higher discount rates put less weight on later years, the preferred option can flip depending on where one stops and how strongly one discounts (\Cref{fig:discount-dynamics}b). A discount rate of 0 makes the model alternate between choices indefinitely (\Cref{fig:discount-dynamics}c), meaning that the decision horizon is effectively infinite. A positive discount (3\%) stabilises the choice if the horizon is sufficiently long at around $\sim2027$, suggesting a decision horizon of ~17 years. Conversely, a negative discount (-1\%) exacerbates the alternation between choices the longer the horizon is.

\clearpage
\section{Review screening phase}
\label{app:screening}

\begin{figure}[hbt]
    \centering
    \begin{subfigure}{0.6\linewidth}
        \includegraphics[width=\linewidth]{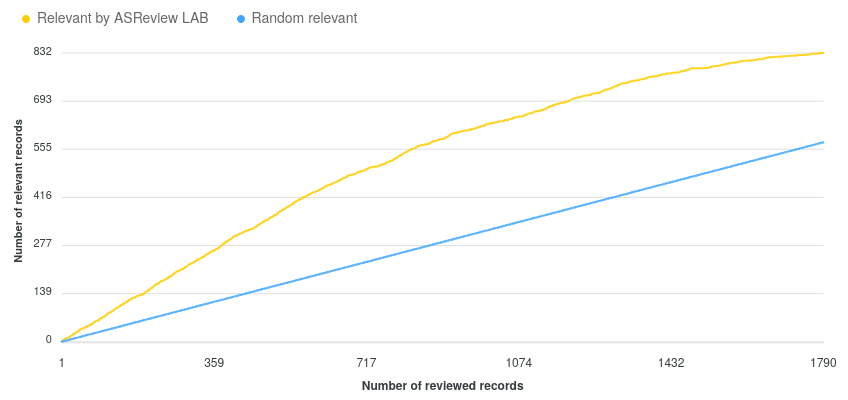} 
        \caption{}
    \end{subfigure}
    \begin{subfigure}{0.6\linewidth}
        \includegraphics[width=\linewidth]{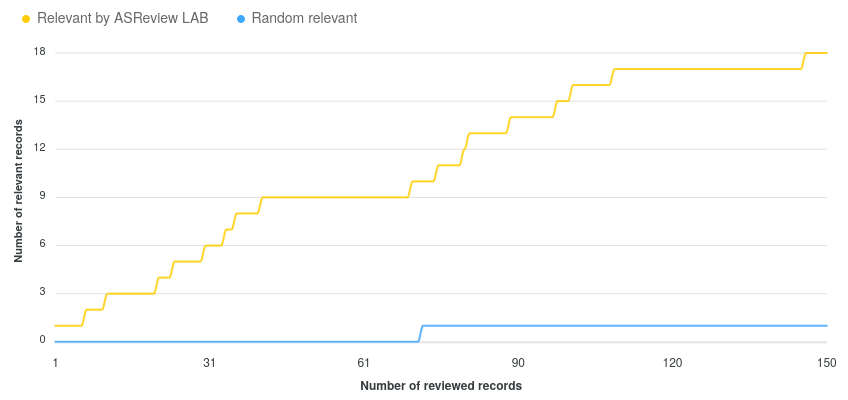} 
    \end{subfigure}
    \begin{subfigure}{0.6\linewidth}
        \includegraphics[width=\linewidth]{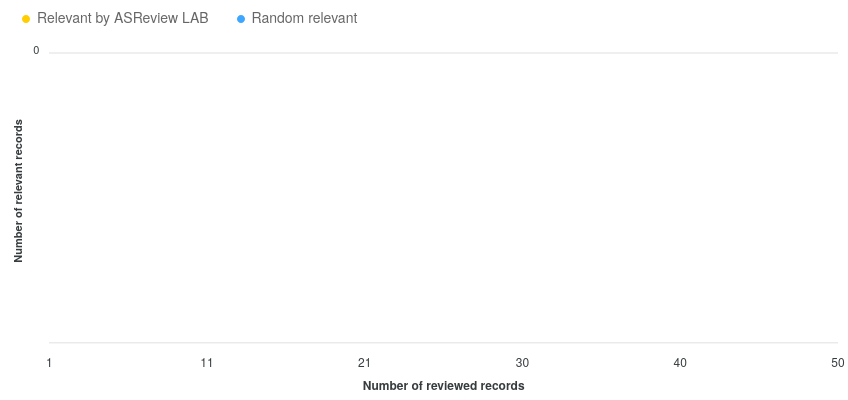} 
    \end{subfigure}
    \caption{Screening phases: relevant records (yellow) versus random relevant (blue). Figures produced using ASReview LAB version 1.6.6~\parencite{developers_asreview_2025}. \textbf{i} using Naive Bayes as classifier and a \gls{td-idf} feature extractor. This phase ended after screening ~66\% of the sample. \textbf{Screening phase ii} using a two-layer neural network as classifier and a Doc2Vec feature extractor. This phase ended after screening ~75\% of the sample. \textbf{Screening phase iii} using a random sample of 50 leftover studies. No relevant records found.}
    \label{fig:screening}
\end{figure}

Screening was subdivided into three steps: (i) an initial pass using the default Naive Bayes classifier with a \gls{td-idf} feature extractor, with a stopping condition of 66\% of total papers screened; (ii) a second pass using a more computationally expensive two layer neural network with Doc2Vec as the feature extractor~\parencite{lau_empirical_2016}, with a stopping condition of 75\% of papers screened; (iii) a final pass where a sample of 50 random leftover records was screened to assess the effectiveness of the active learning tool. The progress at each stage of the screening phase is depicted in \autoref{fig:screening}. It can be seen that, at the end of stage (i), the rate of relevant studies significantly decreases in relation to those analysed meaning that further screening would be mostly immaterial, which was confirmed during stage (ii). None of these random records evaluated at stage (iii) was deemed relevant, which led us to conclude that the screening tool was an effective aid.

\clearpage

\section{Assessment questionnaire}

% Please add the following required packages to your document preamble:
% \usepackage{booktabs}
\begin{table}[!h]
\caption{Questionnaire filled for each study in our sample. Question had a multiple-choice setup with two exceptions: (-) indicates an answer was open text, and (*) indicates an answer could by any natural number.}
\label{tab:questions}
\begin{tabular}{@{}llp{0.33\linewidth}p{0.33\linewidth}@{}}
\toprule
Stage          & Subject             & Question                                                                & Options                                                                                         \\ \midrule
Eligibility    &                     & Is the study relevant at or above national scale?                       & Yes / No / Unclear                                                                              \\ \cmidrule(l){3-4} 
               &                     & Is the study relevant to at least one self-contained ESOM?              & Yes / No / Unclear                                                                              \\ \cmidrule(l){3-4} 
               &                     & Does the study relate to energy generation capacity expansion planning? & Yes / No / Unclear                                                                              \\ \cmidrule(l){3-4} 
               &                     & Does the study relate to pathways?                                      & Yes / No / Unclear                                                                              \\ \midrule
Classification &                     & What type of study is this?                                             & Single model study / Review / Model inter-comparison / Unclear                                  \\ \midrule
Analysis       & Model detail        & What is the name of the modelling framework used?                       & - / Unclear                                                                                     \\ \cmidrule(l){3-4} 
               &                     & What name did the authors use for this specific model?                  & - / Unclear                                                                                     \\ \cmidrule(l){3-4} 
               &                     & What type of formulation was used?                                      & LP / MILP / NLP / MINLP / Other / Unclear                                                       \\ \cmidrule(l){2-4} 
               & Spatial detail      & What region of the world was modelled?                                  & Global / Africa / Asia / Europe / North America / Oceania / South America / Synthetic / Unclear \\ \cmidrule(l){3-4} 
               &                     & How many regions does the model have?                                   & * / Unclear                                                                                     \\ \cmidrule(l){2-4} 
               & Temporal detail     & Which types of foresight were used?                                     & Perfect / Limited / Myopic / Unclear                                        \\ \cmidrule(l){3-4} 
               &                     & What is the baseline year of the model?                                 & * / Unclear                                                                                     \\ \cmidrule(l){3-4} 
               &                     & What is the target year of the exercise?                                & * / Unclear                                                                                     \\ \cmidrule(l){3-4} 
               &                     & What was the final year modelled?                                       & * / Unclear                                                                                     \\ \cmidrule(l){3-4} 
               &                     & How many investment decision years were used?                           & * / Unclear                                                                                     \\ \cmidrule(l){3-4} 
               &                     & How many operational timesteps were used?                               & * / Unclear                                                                                     \\ \cmidrule(l){2-4} 
               & Investment dynamics & What type of technological learning was used?                           & Exogenous / Endogenous / Unclear                                                                \\ \bottomrule
\end{tabular}
\end{table}

\clearpage
\section{Regions used for spatial classification}

% Please add the following required packages to your document preamble:
% \usepackage{booktabs}
\begin{table}[!h]
\centering
\caption{World region classification used in this study as defined in Our World in Data~\parencite{our_world_in_data_team_definitions_2018}. Please note that a country being in this list does not imply it was individually featured within in our sample of single-model studies.}
\begin{tabular}{@{}p{0.2\linewidth}p{0.79\linewidth}@{}}
\toprule
\textbf{Region} & \textbf{Countries} \\ \midrule
Africa & Algeria, Angola, Benin, Botswana, Burkina Faso, Burundi, Cameroon, Cape Verde, Central African Republic, Chad, Comoros, Congo, Cote d'Ivoire, Democratic Republic of Congo, Djibouti, Egypt, Equatorial Guinea, Eritrea, Eswatini, Ethiopia, Ethiopia (former), Gabon, Gambia, Ghana, Guinea, Guinea-Bissau, Kenya, Lesotho, Liberia, Libya, Madagascar, Malawi, Mali, Mauritania, Mauritius, Mayotte, Morocco, Mozambique, Namibia, Niger, Nigeria, Orange Free State, Reunion, Rwanda, Saint Helena, Sao Tome and Principe, Senegal, Seychelles, Sierra Leone, Somalia, South Africa, South Sudan, Sudan, Sudan (former), Tanzania, Togo, Tunisia, Uganda, Western Sahara, Zambia, Zimbabwe \\  \hline
Asia & Afghanistan, Armenia, Azerbaijan, Bahrain, Bangladesh, Bhutan, British Indian Ocean Territory, Brunei, Cambodia, China, Christmas Island, Cocos Islands, Democratic Republic of Vietnam, East Timor, Georgia, Hong Kong, India, Indonesia, Iran, Iraq, Israel, Japan, Jordan, Kazakhstan, Korea (former), Kuwait, Kyrgyzstan, Laos, Lebanon, Macao, Malaysia, Maldives, Mongolia, Myanmar, Nepal, North Korea, Oman, Pakistan, Pakistan (former), Palestine, Philippines, Qatar, Republic of Vietnam, Saudi Arabia, Singapore, South Korea, Sri Lanka, Syria, Taiwan, Tajikistan, Thailand, Turkey, Turkmenistan, United Arab Emirates, Uzbekistan, Vietnam, Yemen, Yemen Arab Republic, Yemen People's Republic \\  \hline
Europe & Aland Islands, Albania, Andorra, Austria, Austria-Hungary, Belarus, Belgium, Bosnia and Herzegovina, Bulgaria, Croatia, Cyprus, Czechia, Czechoslovakia, Denmark, Duchy of Modena and Reggio, Duchy of Parma and Piacenza, East Germany, Estonia, Faroe Islands, Finland, France, Germany, Gibraltar, Grand Duchy of Baden, Grand Duchy of Tuscany, Greece, Guernsey, Hungary, Iceland, Ireland, Isle of Man, Italy, Jersey, Kingdom of Bavaria, Kingdom of Sardinia, Kingdom of Saxony, Kingdom of Wurttemberg, Kingdom of the Two Sicilies, Kosovo, Latvia, Liechtenstein, Lithuania, Luxembourg, Malta, Moldova, Monaco, Montenegro, Netherlands, North Macedonia, Norway, Poland, Portugal, Romania, Russia, San Marino, Serbia, Serbia and Montenegro, Slovakia, Slovenia, Spain, Sweden, Switzerland, USSR, Ukraine, United Kingdom, Vatican, West Germany, Yugoslavia \\  \hline
North America & Anguilla, Antigua and Barbuda, Aruba, Bahamas, Barbados, Belize, Bermuda, Bonaire Sint Eustatius and Saba, British Virgin Islands, Canada, Cayman Islands, Costa Rica, Cuba, Curacao, Dominica, Dominican Republic, El Salvador, Federal Republic of Central America, Greenland, Grenada, Guadeloupe, Guatemala, Haiti, Honduras, Jamaica, Martinique, Mexico, Montserrat, Netherlands Antilles, Nicaragua, Panama, Puerto Rico, Saint Barthelemy, Saint Kitts and Nevis, Saint Lucia, Saint Martin (French part), Saint Pierre and Miquelon, Saint Vincent and the Grenadines, Sint Maarten (Dutch part), Trinidad and Tobago, Turks and Caicos Islands, United States, United States Virgin Islands \\  \hline
Oceania & American Samoa, Australia, Cook Islands, Fiji, French Polynesia, Guam, Kiribati, Marshall Islands, Micronesia (country), Nauru, New Caledonia, New Zealand, Niue, Norfolk Island, Northern Mariana Islands, Palau, Papua New Guinea, Pitcairn, Samoa, Solomon Islands, Tokelau, Tonga, Tuvalu, Vanuatu, Wallis and Futuna \\  \hline
South America & Argentina, Bolivia, Brazil, Chile, Colombia, Ecuador, Falkland Islands, French Guiana, Great Colombia, Guyana, Paraguay, Peru, South Georgia and the South Sandwich Islands, Suriname, Uruguay, Venezuela \\ \bottomrule
\end{tabular}
\label{tab:world-regions}
\end{table}

\clearpage
\section{Bibliometric analyses}

Additional bibliometric analyses were conducted using the pybibx library to analyse our sample~\parencite{pereira_pybibx_2025}. To do so, bibliometric data for each study was downloaded from the SCOPUS database~\parencite{elsevier_bv_scopus_2025} and subsequently fed to pybibx library~\parencite{pereira_pybibx_2025} for further processing and cleaning. In the case of institutional affiliation analyses (\cref{tab:co-authorship-regions,tab:co-authorship-countries}) a total of five cases---0.3\% of the sample---could not be assigned to a country, meaning the total co-authorship sample including unknowns was 1450, meaning missing affiliation cases are immaterial to our results.

\begin{figure}[hbt]
    \centering
    \includegraphics[width=1\linewidth]{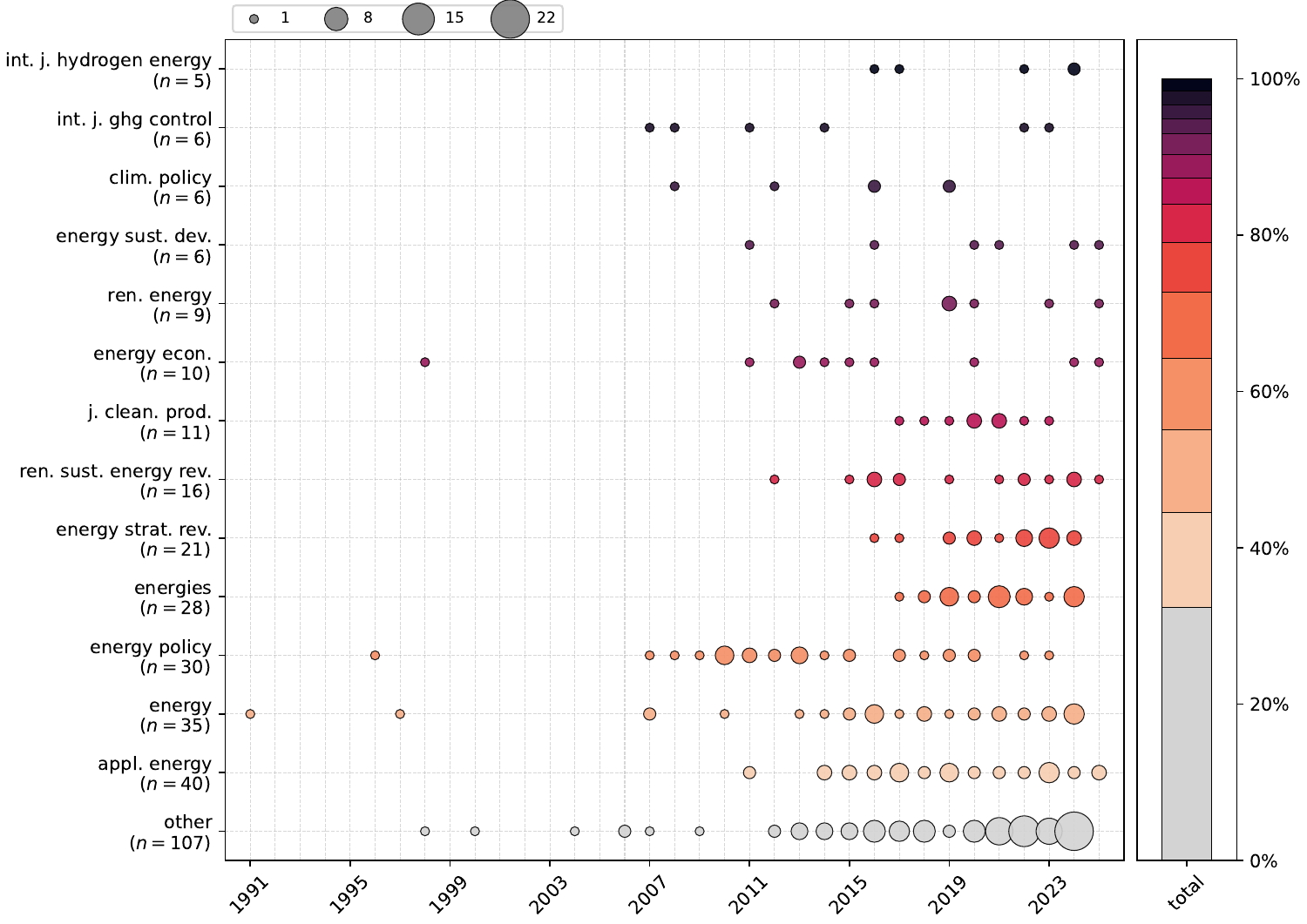}
    \caption{Publishing trends in our full sample of studies ($n=330$). Size depicts study count per year. Journals with less than 5 total publications were aggregated into ``other''.}
    \label{fig:biblio-journal}. 
\end{figure}

% Please add the following required packages to your document preamble:
% \usepackage{booktabs}
\begin{table}[hbt]
\centering
\caption{Co-authorship share by institution for 1445 co-authors identified within our full sample of studies ($n=330$) for each world region as defined by Our World in Data~\parencite{our_world_in_data_team_definitions_2018}.}
\label{tab:co-authorship-regions}
\begin{tabular}{@{}lp{0.4\linewidth}lll@{}}
\toprule
World region  & Countries identified                                                                                                                                                                                                                                  & N. of countries & Frequency & Share \\ \midrule
Africa        & Ethiopia, Ghana, Morocco, Tunisia, Egypt, South Africa, Tanzania, Kenya, Namibia                                                                                                                                                                      & 9               & 31                   & 2.1\%          \\ \hline
Asia          & China, Japan, South Korea, India, Thailand, Iran, Indonesia, Taiwan, Singapore, Syria, Pakistan, Turkey, Bangladesh, Jordan, Nepal, Kazakhstan, Saudi Arabia, Sri Lanka                                                                               & 18              & 283                  & 19.6\%         \\ \hline
Europe        & Germany, United Kingdom, Netherlands, Italy, Ireland, Finland, Denmark, Sweden, Norway, Austria, France, Spain, Switzerland, Portugal, Greece, Cyprus, Poland, Belgium, Iceland, Ukraine, Czechia, Faroe Islands, Russia, Estonia, Latvia, Luxembourg & 26              & 839                  & 58.1\%         \\ \hline
North America & United States, Canada, Mexico, Costa Rica                                                                                                                                                                                                             & 4               & 197                  & 13.6\%         \\ \hline
Oceania       & Australia, New Zealand                                                                                                                                                                                                                                & 2               & 16                   & 1.1\%          \\ \hline
South America & Brazil, Chile, Bolivia, Colombia, Ecuador                                                                                                                                                                                                             & 5               & 79                   & 5.5\%          \\ \bottomrule
\end{tabular}
\end{table}

% Please add the following required packages to your document preamble:
% \usepackage{booktabs}
\begin{table}[hbt]
\centering
\caption{Countries with the largest co-authorship share by institutional affiliation. Computed using pybibx~\parencite{pereira_pybibx_2025}.}
\label{tab:co-authorship-countries}
\begin{tabular}{@{}lll@{}}
\toprule
Country & Frequency & Share ($n=1445$)\\ \midrule
Germany & 176 & 12.2\% \\
United States of America & 132 & 9.1\%  \\
United Kingdom & 129 & 8.9\%  \\
China & 102 & 7.1\%  \\
Netherlands & 91  & 6.3\%  \\
Italy & 80  & 5.5\%  \\
Japan & 54  & 3.7\%  \\
Ireland & 48  & 3.3\%  \\
Finland & 42  & 2.9\%  \\
Denmark & 39  & 2.7\%  \\
Sweden & 39  & 2.7\%  \\
Canada & 39  & 2.7\%  \\
Norway & 36  & 2.5\%  \\
South Korea & 35  & 2.4\%  \\
Austria & 31  & 2.1\%  \\
Brazil & 28  & 1.9\%  \\
Chile & 26  & 1.8\%  \\
France & 25  & 1.7\%  \\
Spain & 22  & 1.5\%  \\
Switzerland & 21  & 1.5\%  \\
India & 19  & 1.3\%  \\
Portugal & 16  & 1.1\%  \\
Mexico & 16  & 1.1\%  \\
Thailand & 16  & 1.1\%  \\
Iran & 15  & 1.0\%  \\
Indonesia & 13  & 0.9\%  \\
Greece & 12  & 0.8\%  \\
Australia & 12  & 0.8\%  \\
Costa Rica & 10  & 0.7\%  \\
Bolivia & 9   & 0.6\%  \\ \bottomrule
\end{tabular}
\end{table}

\clearpage
\section{Intercoder analysis}

We conducted inter-coder analyses for ten randomised article within our sample of single model studies ($n=305$) to evaluate causes of disagreement between our two coders, with each coder evaluating five from the other. We subdivided evaluations into qualitative questions and quantitative questions.

For qualitative questions we made use of Cohen's kappa ($\kappa$), as shown in \cref{tab:intercoder-qualitative}. Four out of seven questions showed perfect agreement, and the remaining three showed comparatively poor agreement. An additional control test, $\kappa_c$, was conducted by removing cases where either reviewer marked the question as indeterminate, which effectively removed all disagreement.

\Cref{tab:intercoder-quantitative} shows the computation of Pearson's $r$ for our quantitative questions, with broad agreement in all categories except for \textbf{Q17}. Disagreement was mostly caused by a study that used ambiguous language when stating the horizon of the exercise~\parencite{limmeechokchai_energy_2022}, which lead to coders noting different milestones based either on textual descriptions or the figures shown in the article (35 and 7, respectively).

\begin{table}[hbt]
\caption{Inter-coder analysis of qualitative questions conducted for ten single model studies. $\kappa$ is Cohen's kappa, $p_o$ is the observed agreement among raters, $p_e$ is the hypothetical probability of chance agreement, $\kappa_c$ is a re-computation of Cohen's kappa with indeterminate cases removed. Cases with a single labelled category ($p_e = 1$) are marked as ``inf''.}
\label{tab:intercoder-qualitative}
\centering
\begin{tabular}{@{}lrrrr@{}}
\toprule
Question              & $\kappa$    & $p_o$  & $p_e$   & $\kappa_c$ \\ \midrule
Study type            & inf         & 1      & 1       & inf        \\
Modelling framework   & 1           & 1      & 0.28    & 1          \\
Model name            & 1           & 1      & 0.52    & 1          \\
Formulation           & 0.44        & 0.8    & 0.64    & 1          \\
World region          & 1           & 1      & 0.26    & 1          \\
Foresight             & 0.23        & 0.6    & 0.48    & 1          \\
Endogenous learning   & 0.41        & 0.8    & 0.66    & 1          \\ \bottomrule
\end{tabular}%
\end{table}

\begin{table}[hbt]
\centering
\caption{Inter-coder analysis of quantitative questions conducted for ten single model studies. $r$ is Pearson's r, $p$ is the p-value. Indeterminate cases were removed during calculations.}
\label{tab:intercoder-quantitative}
\begin{tabular}{@{}lrrr@{}}
\toprule
Question             & $r$        & $p$       & Indeterminate count \\ \midrule
Number of regions    & 1          & 0         & 4                   \\
Baseline year        & 0.96       & 8.19e-06  & 0                   \\
Target year          & 1          & 0         & 0                   \\
Horizon year         & 0.94       & 0.0002    & 1                   \\
Number of milestones & 0.57       & 0.087     & 0                   \\
Number of timesteps  & 0.999      & 1.07e-06  & 4                   \\ \bottomrule
\end{tabular}%
\end{table}

\clearpage

\section{Model resolution analysis}

\begin{figure}[!hbt]
    \centering
    \includegraphics[width=0.7\linewidth]{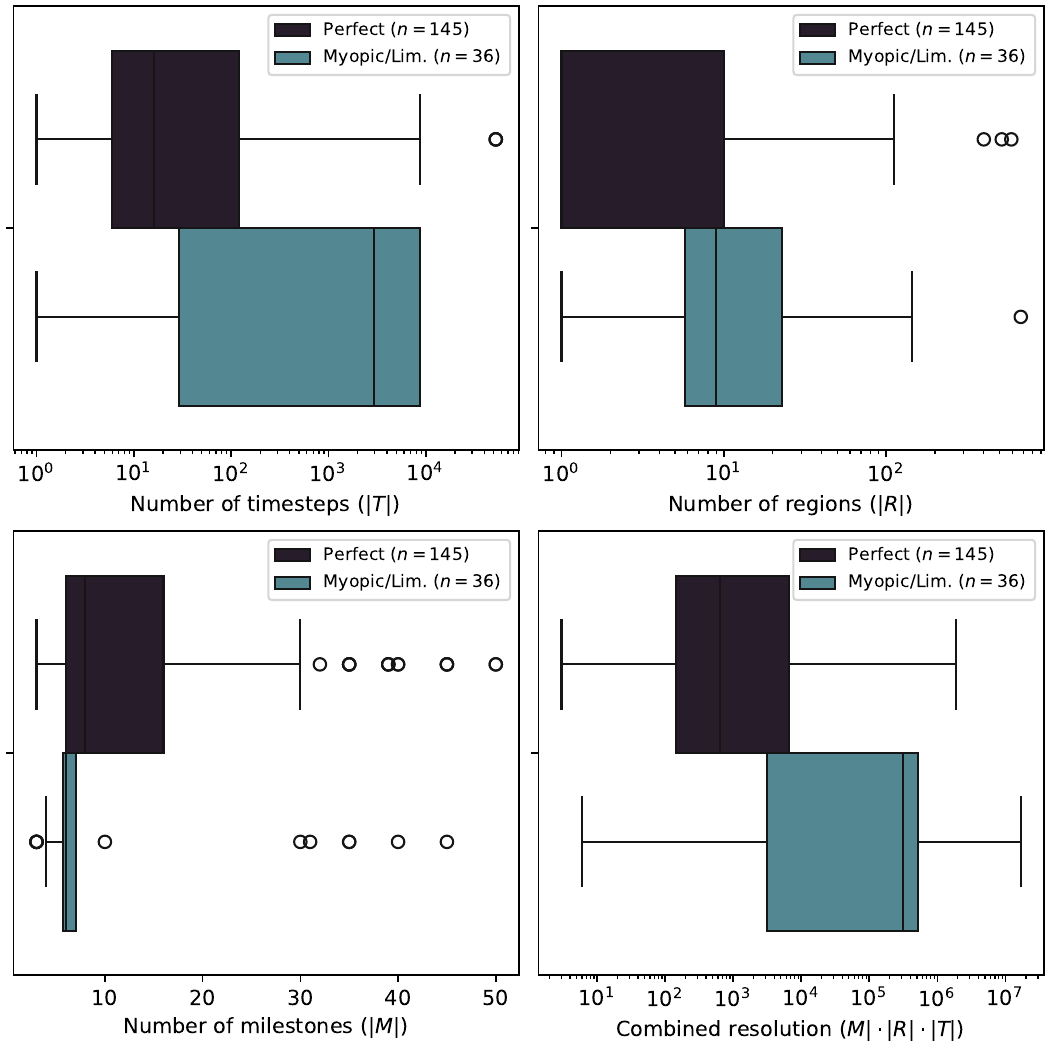}
    \caption{Aggregated dimensional characteristics of single model studies with sufficient transparency ($n=181$). Combined resolution ($|M| \cdot |R| \cdot |T|$) does not necessarily represent computational burden in myopic or limited foresight models given that they are rolling horizon approaches.}
    \label{fig:resolution-aggregate}
\end{figure}

\begin{figure}[!hbt]
    \centering
    \includegraphics[width=0.9\linewidth]{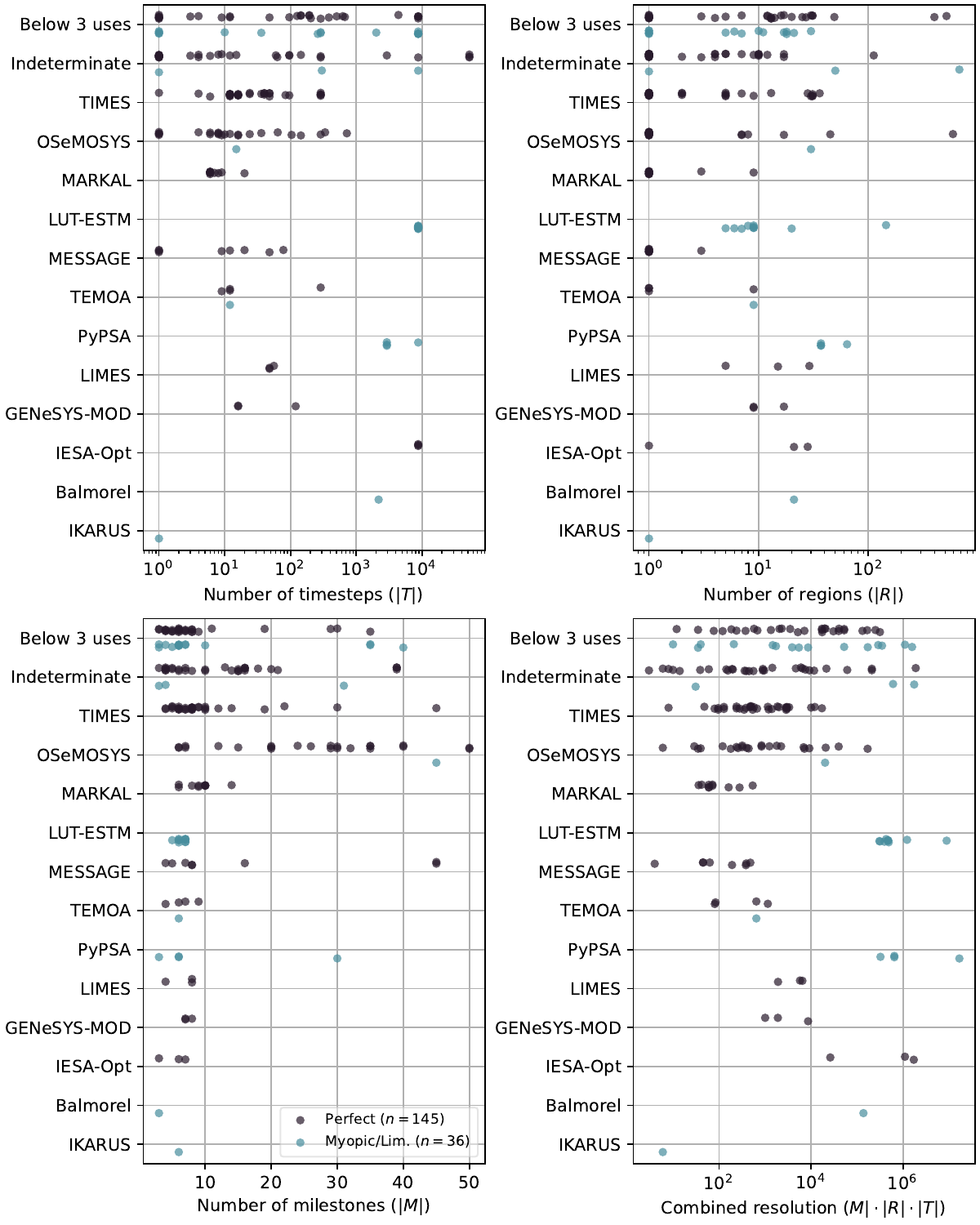}
    \caption{Summary of dimensional characteristics of single model studies with sufficient transparency by model framework ($n=181$). Combined resolution ($|M| \cdot |R| \cdot |T|$) does not represent computational burden in myopic or limited foresight models given that they are rolling horizon approaches.}
    \label{fig:resolution-by-framework}
\end{figure}

\clearpage
\section{Long-term heuristics}
% Please add the following required packages to your document preamble:
% \usepackage{booktabs}
\begin{table}[hbt]
\centering
\caption{Modelling frameworks with detected end-of-horizon extensions in our sample if single-model studies with sufficient transparency ($n=181$).}
\label{tab:eoh-studies}
\begin{tabular}{@{}lrp{0.6\linewidth}@{}}
\toprule
Framework & Counts & References \\ \midrule
DIMENSION       & 1      & \textcite{jagemann_decarbonizing_2013}           \\
GridPath        & 1      & \textcite{chowdhury_enabling_2022}           \\
IKARUS          & 1      & \textcite{martinsen_compromises_2008}           \\
MARKAL          & 1      &  \textcite{kannan_development_2011}          \\
MESSAGE          & 1      &  \textcite{ghadaksaz_energy_2020}          \\
OSeMOSYS        & 6      &  \textcite{slimani_towards_2024,fernandez_vazquez_analyzing_2022,dallmann_between_2022, henke_exploring_2024,moksnes_increasing_2024,taliotis_indicative_2016}          \\
SWITCH          & 1      & \textcite{verastegui_optimization-based_2021}           \\
TIMES           & 3      & \textcite{pattupara_alternative_2016, carvajal_large_2019, sandberg_impact_2022}           \\ \bottomrule
\end{tabular}
\end{table}

\clearpage
\printbibliography
\end{refsection}

\end{document}